\documentclass{tlp}
\usepackage{aopmath}
\usepackage{graphicx}
\usepackage{latexsym}
\usepackage{amssymb}

\newcommand{\nop}[1]{}
\newtheorem{definition}{Definition}
\newtheorem{example}{Example}

\newcommand{\OPT}{\mbox{$ O\!P\!T$}}

\newcommand{\DA}{\mbox{\tt DATALOG}}
\newcommand{\DAm}{\mbox{\DA$^\neg$}}
\newcommand{\DAs}{\mbox{\DA$^{\neg_s}$}}
\newcommand{\DAd}{\mbox{\DA$^{\oplus}$}}
\newcommand{\DAc}{\mbox{\DA$^{\small \Leftarrow}$}}
\newcommand{\DAsc}{\mbox{\DA$^{\neg_s,{\small \Leftarrow}}$}}
\newcommand{\DAdc}{\mbox{\DA$^{\oplus,{\small \Leftarrow}}$}}
\newcommand{\DAsd}{\mbox{\DA$^{\neg_s,\oplus}$}}
\newcommand{\DAsdc}{\mbox{\DA$^{\neg_s,\oplus,{\small\Leftarrow}}$}}
\newcommand{\NPDA}{\mbox{${\cal N\!P\, D}atalog$}}

\newcommand{\bldl}{\smallskip\[\tt\begin{array}{ll}}
\newcommand{\cldl}{\vspace{-0.4cm}\[\tt\begin{array}{ll}}
\newcommand{\eldl}{\end{array}\]\rm}

\newcommand{\XX}{\mbox{\bf X}}
\newcommand{\YY}{\mbox{\bf Y}}

\def\punto {\hspace*{\fill}$\Box$}
\newcommand{\items}[1]{\vspace*{#1 mm}\item}

\newcommand{\TD}   {\mbox{$\cal T_D$}}
\newcommand{\TQ}   {\mbox{$\cal T_Q$}}
\newcommand{\TP}   {\mbox{$\cal T_P$}}
\newcommand{\WQ}   {\mbox{$\cal W_Q$}}
\newcommand{\WP}   {\mbox{$\cal W_P$}}

\newcommand{\LL}   {\mbox{$\cal L$}}

\newcommand{\NQ}  {\mbox{$\cal N\!Q$}}
\newcommand{\NP} {\mbox{$\cal N\!P$}}
\newcommand{\PT} {\mbox{$\cal PTIME$}}

\newcommand{\coNP}{\mbox{$co\NP$}}

\newcommand{\NPMV} {{\mbox{$\cal N\!P\!M\!V$}}}
\newcommand{\npmv} {{\mbox{$\cal N\!P\!M\!V$}}}
\newcommand{\NQPMV} {{\mbox{$\cal Q\!N\!P\!M\!V$}}}
\newcommand{\coNPMV} {{\mbox{$co\NPMV$}}}
\newcommand{\conpmv} {{\mbox{$co\NPMV$}}}
\newcommand{\coNQPMV} {{\mbox{${\cal Q}co\NPMV$}}}

\def\PP{{\cal P}}

\def\MM{{\cal MM}}
\def\SM{{\cal SM}}
\def\QQ{{\cal Q}}
\def\SS{{\cal S}}

\newcommand{\DB} {\mbox{$\cal DB$}}
\newcommand{\DS} {\mbox{$\cal DS$}}
\newcommand{\D} {\mbox{$\cal DB$}}

\def\<{\langle}
\def\>{\rangle}
\def\+{\mbox{+}}
\def\-{\mbox{-}}
\def\={\mbox{\rm =}}
\def\.{\mbox{\rm .}}
\def\({\begin{center}$}
\def\){$\end{center}}
\def\"{``}

\newcommand{\vs}{\vspace*{-0mm}}
 \submitted {4 April 2008}
 \revised {2 September 2009}
 \accepted{2 November 2009}

\begin{document}
\bibliographystyle{acmtrans}

\date{}
\title[\NPDA]{\Large $\NPDA$: a Logic Language for Expressing \NP Search and Optimization Problems}

\author[Greco, Molinaro, Trubitsyna, Zumpano]
{Sergio Greco, Cristian Molinaro, Irina Trubitsyna and Ester Zumpano \\ \ \\
DEIS \\
Universit\`a della Calabria \\
87036 Rende, Italy \\
\{greco, cmolinaro, irina, zumpano\}@deis.unical.it}

\pagerange{\pageref{firstpage}--\pageref{lastpage}}

\setcounter{page}{1}

\maketitle

\begin{abstract}
This paper presents a logic language for expressing \NP\ search and
optimization problems.
Specifically, first a language obtained by extending (positive) \DA\
with  intuitive and efficient constructs (namely, stratified negation,
constraints and exclusive disjunction) is introduced.
Next, a further restricted language only using a restricted
form of disjunction to define (non-deterministically) subsets (or
partitions) of relations is investigated.
This language, called \NPDA, captures the power of \DAm\ in expressing search and
optimization problems.
A system prototype implementing \NPDA\ is presented.
The system translates \NPDA\ queries into OPL programs which are executed
by the ILOG OPL Development Studio.
Our proposal combines easy formulation of problems, expressed by means of a
declarative logic language, with the efficiency of the ILOG System.
Several experiments show the effectiveness of this approach.
\end{abstract}


\section{Introduction}

It is well-known that \NP\ search problems can be formulated by means of \DAm\
(Datalog with unstratified negation) queries under non-deterministic stable model semantics so that each
stable model corresponds to a possible solution
\cite{MarTru91,Sac94b}. \NP\ optimization problems can be
formulated by adding a \emph{max} (or \emph{min}) construct to
select the stable model (thus, the solution) which maximizes
(resp., minimizes) the result of a polynomial function applied to
the answer relation. For instance, consider the {\em Vertex Cover} problem of the following example.

\begin{example}\label{Example1-Intro}
Given an undirected graph $G = \<N,E\>$, a subset $V$ of the vertexes
$N$ is a \emph{vertex cover} of $G$  if every edge of $G$
has at least one end in $V$.
The problem can be formulated in terms of
the query $\<\tt \PP_{\ref{Example1-Intro}},v(X)\>$, where $\tt \PP_{\ref{Example1-Intro}}$ is the following $\DAm$ program:
\[
\begin{array}{l}
\tt v(X) \leftarrow node(X), \ \neg nv(X). \\
\tt nv(X) \leftarrow node(X), \ \neg v(X). \\
\tt c \leftarrow edge(X,Y), \neg v(X), \neg v(Y), \ \neg c.
\end{array}
\]
and the predicates $\tt node$ and $\tt edge$ define, respectively, the vertexes
and the edges of the graph by means of a suitable number of facts.
The first two rules define a partition of the relation $\tt node$ ($\tt v$ being the vertex cover),
whereas the last one enforces every stable model to correspond
to some vertex cover as it is satisfied only if the conjunction
$\tt edge(X,Y), \neg v(X), \neg v(Y)$ is false
(otherwise the program does not have stable models).

The min vertex cover problem can be expressed by selecting a
stable model which minimizes the number of elements in $\tt v$;
this is expressed by means of the query
$\<\tt \PP_{\ref{Example1-Intro}},\tt min|v(X)|\>$.~\hfill $\Box$
\end{example}

The problem in using \DAm\ to express search and optimization
problems is that the use of unrestricted negation is often neither
simple nor intuitive and besides it does not allow the expressive
power and complexity of queries to be limited.
For instance, in the example above, the use of explicit constraints instead
of standard rules would permit the distinction between rules used to infer true atoms and rules used
to check properties to be satisfied.


In this paper, in order to enable a simpler and more
intuitive formulation for search and optimization problems and an
efficient computation of queries, \DA-like languages extending
positive \DA\ with intuitive and efficient constructs are
considered.
The first language we present, denoted by \DAsdc, extends the
simple and intuitive structure of \DAs\ (\DA\ with stratified
negation~\cite{Ull88}) with two other types of `controlled'
negation: rules with exclusive {\em
disjunctive} heads and {\em constraint} rules.
The same expressive power as \DAm\ is achieved by such a language.
Next, we propose a further restricted language, called
\emph{\NPDA}, where head disjunction is only used to define
(non-deterministically) partitions of relations. This language
allows us to express, in a simple and intuitive way, both \NP\
search and optimization problems.
As an example, let us consider again the Vertex Cover problem.

\begin{example}\label{Vertex-Cover}
The search query of the previous example can be expressed as
$\<\PP_2,{\tt v(X)}\>$ with $\PP_2$ defined as follows:
\[
\begin{array}{l}
\tt v(X) \oplus nv(X) \leftarrow node(X). \hspace{2cm} \\
\tt \Leftarrow edge(X,Y), \neg v(X), \neg v(Y).
\end{array}
\]
where $\oplus$ denotes exclusive disjunction, i.e., if the body
of the rule is true, then exactly one atom in the head is true.
The rule with empty head defines a constraint, i.e.,
a rule which is satisfied only if the body is false.
The first rule guesses a partition of $\tt node$ whereas the second one
is a constraint stating that two
connected nodes cannot be both outside  the cover, which is defined by the
nodes belonging to $\tt v$. \hfill $\Box$
\end{example}

\paragraph{Contribution.} \ \\
The main contribution of this paper is
the proposal of a simple and intuitive language where the use of
stable model semantics allows us to refrain from uncontrolled
forms of unstratified negation\footnote{The constructs here
considered, essentially, force the use of a restricted form of
unstratified negation.} and avoid both undefinedness and
unnecessary computational complexity.

More precisely, the paper  presents the language \emph{\NPDA},
which extends \DAs\ with constraints and head disjunction,
where the latter is used only to define (non-deterministically)
partitions of ``deterministic'' relations. This language allows
both \NP\ search and optimization problems to be expressed in a
simple and intuitive way.

The simplicity of the \NPDA\ language  enables queries to be
easily translated into other formalisms such as constraint
programming languages, which are well-suited to compute programs
defining \NP\ problems. This paper  also shows how \NPDA\
queries can be translated into \emph{OPL (Optimization Programming
Language)} \cite{VAN99,VAN99b} programs.

Several examples of queries expressing \NP\ problems
suggest that logic formalisms allow an easy formulation of
queries. On the other hand, constraint programming systems permit an efficient execution.
Therefore, \NPDA\ can also be used to define a logic interface for
constraint programming solvers.
We have implemented a system prototype which translates \NPDA\ queries
into OPL programs, which are then executed by means of
the ILOG OPL Development Studio  \cite{Ilog}.
The effectiveness of our approach is demonstrated by several
experiments comparing \NPDA\ with other systems.

With respect to other logic languages previously proposed
\cite{CadIan01,CadIan05,leone-sys,GreSac*95,smodels}, the novelty of the
paper is that it considers an answer set programming language able to
express the complete set of \NP\ decision, search and optimization
problems, by using a restricted form of unstratified negation.

\paragraph{Organization.}
The paper is organized as follows.
Section~2 introduces syntax and semantics of \DAm
and its ability to express \NP\ search and optimization queries under
non-deterministic stable model semantics.
Section~3 introduces the \DAsdc\ language, and shows its ability to express
\NP\ search and optimization problems.
Section~4 presents the \NPDA\ language, obtained introducing
simple restrictions to \DAsdc\ and shows that \NPDA\ has the
same expressive power as \DAsdc\ and \DAm.
Section~5 illustrates how \NPDA\
queries  can be translated into OPL programs and presents several experiments
showing the effectiveness of the proposed approach.
Section~6 discusses
several related languages and systems recently proposed in the
literature.
Finally, conclusions are drawn in Section~7.

\section{\DAm}


\noindent It is assumed that the reader is familiar with the basic
terminology and notation of relational databases and database
queries \cite{AHV94,Ull88}.

\paragraph{Syntax.}

A \DAm\ rule $r$ is of the form $A \leftarrow
B_1,\dots,B_m, \neg B_{m+1}, \dots, \neg B_n$, where $A$ is an
atom ({\em head} of the rule) and $B_1,\dots,B_m, \neg B_{m+1},
\dots, \neg B_n$ (with $n \geq 0$) is a conjunction of literals
({\em body} of the rule). A fact is a ground rule with empty body.
Generally, predicate symbols are partitioned into two different
classes: extensional (or EDB), i.e. defined by the ground facts of
a database, and intensional (or IDB), i.e. defined by the rules of
the program. The definition of a predicate $p$ consists of all the
rules (or facts) having $p$ in the head.

A database $D$ consists of all the facts defining EDB predicates,
whereas a \DAm\ program $\PP$ consists of the rules defining IDB
predicates.
It is assumed that programs are \emph{safe}
\cite{Ull88}, i.e. variables appearing in the head or in negative
body literals are range restricted as they appear in some positive
body literal, and that possible constants in $\PP$ are taken from
the database domain. For each rule, variables appearing in the
head are said to be universally quantified, whereas the remaining
variables are said to be existentially quantified.

The class of all \DAm\ programs is simply called \DAm; the
subclass of all positive (resp. stratified) programs is called
\DA\ (resp. \DAs) \cite{AHV94}. Observe that $\DA\ \subseteq \DAs\
\subseteq \DAm$ (the class of Datalog queries with possibly
unstratified negation).

\paragraph{Semantics.}

The semantics of a positive program $\PP$ is given by the unique
minimal model $\MM(\PP)$.
The semantics of programs with negation $\PP$ is given by the set
of its stable models $\SM(\PP)$. An interpretation $M$ is a
\emph{stable model} (or \emph{answer set}) of $\PP$ if $M$ is the
unique minimal model of the positive program $\PP^M$, where $\PP^M$
denotes the positive logic program obtained from $ground(\PP)$ by removing
(i) all rules $r$ such that  there is a negative literal
$\neg A$ in the body of $r$ and $A$ is in $M$, and
(ii) all the negative literals from the remaining rules
\cite{GelLif88}. It is well-known that a program may have $n$
stable models with $n \geq 0$. Stratified programs have a unique
stable model which coincides with the {\em perfect model},
obtained by partitioning the program into an ordered number of
suitable subprograms (called `strata') and computing the fixpoints
of every stratum in their order \cite{Ull88}.
Given a set of ground atoms $S$ and an atom $g(t)$, $S[g]$ (resp.
$S[g(t)]$) denotes the set of $g$-tuples (resp. tuples matching
$g(t)$) in $S$.



\paragraph{\DAm\ Search and Optimization Queries.}

\noindent Search and optimization problems can be expressed using
different logic formalisms such as Datalog with unstratified
negation.

\begin{definition}\label{def-search-query}
A \DAm\ {\em search query} is a pair $Q=\<\PP,g(t)\>$, where
$\PP$ is a \DAm\ program and $g(t)$ is an atom s.t. $g$ is an IDB predicate of $\PP$.
The answer to $Q$ over a database $D$ is $Q(D) =
\{M[g(t)]|\,M \in \SM(\PP \cup D)\}$.
The answer to the \DAm\ {\em optimization query} $opt(Q) = \<\PP,opt|g(t)|\>$,
where $opt$ is either $max$ or $min$, over a database $D$, consists of the
answers in $Q(D)$ with the maximum or minimum (resp., if $opt = max$ or $min$)
cardinality and is denoted by $opt(Q)(D)$.\punto
\end{definition}

\noindent
Observe that, for the sake of simplicity,
optimization queries computing the maximum or minimum cardinality
of the output relation are considered,
although any polynomial function might be used.
Therefore, the answer here considered is a set of sets
of atoms. Possible and certain answers can be obtained by
considering the union or the intersection of the sets, respectively.
Instead of considering possible and certain reasoning, we
introduce non-deterministic answers as follows.

\begin{definition}\label{def-nondeterministic-search-query}
A (\emph{non-deterministic}) answer to a \DAm\ search query $Q$ applied to
a database $D$ is $Q(D) = S$ where $S$ is a relation selected non-deterministically from $Q(D)$.
A (non-deterministic) answer to a
\DAm\ optimization query $opt(Q)$ over a database $D$ is
$opt(Q)(D) = S$ where $S$ is a relation selected
non-deterministically from $opt(Q)(D)$. \punto
\end{definition}

It is worth noting that, like for search queries, also for
optimization queries  the relation with  optimal cardinality
rather than just the cardinality is returned. Thus, given a search
query $Q = \<\PP,g(t)\>$ and a database $D$, the output relation
$Q(D)$ consists of all tuples $g(u)$ matching $g(t)$ and belonging
to a stable model $M$ of $\PP \cup D$, selected
non-deterministically. For a given optimization query $OQ =
\<\PP,min|g(t)|\>$ (resp. $\<\PP,max|g(t)|\>$), the output
relation $OQ(D)$ consists of the set of tuples $g(u)$ matching
$g(t)$ and belonging to a stable model $M$ of $\PP \cup D$, selected
non-deterministically among those which minimize (resp. maximize)
the cardinality of the output relation.
From now on, we concentrate our attention on
non-deterministic queries. An example of a non-deterministic \DAm\
search query is shown in \ Example \ref{Example1-Intro}; the
optimization problem is expressed by rewriting the query goal as
$\<\PP_1,{\tt min|v(X)|}\>$ whose meaning is to further restrict the
set of stable models to those for which $\tt v$ has minimum cardinality.
\nop{+++++++++++++++++++++++++++++++++++++++++++++++++++++++
The set of all \DAm, \DA\
or \DAs\ queries are denoted respectively by $\bf Q^\neg$,  $\bf
Q$ and $\bf Q^{\neg_s}$. Given a class of queries $\cal Q$, the
corresponding sets of search and optimization queries are denoted,
respectively by $search({\cal Q})$ and $opt({\cal Q})$. Thus,
$search({\bf Q}^\neg)$ (resp. $search({\bf Q})$, $search({\bf
Q}^{\neg_s})$) denotes the set of all \DAm\ (resp. \DA, \DAs)
search queries, whereas $opt({\bf Q}^\neg)$ (resp. $opt({\bf Q})$,
$opt({\bf Q^{\neg_s}})$) denotes the set of all \DAm\ (resp. \DA,
\DAs) optimization queries. Observe that, given a database $D$, if
the program $\PP \cup D$ has no stable models, then both search and
optimization queries are not defined on $D$.
+++++++++++++++++++++++++++++++++++++++++++++++++++++++++++++}

In \cite{Sac94b} and \cite{GreSac97} it has been shown that \DAm\ search and optimization queries
under (non-deterministic) stable model semantics express the class
of $\NP$ search and optimization problems (denoted, respectively, by $\NQPMV$ and $\OPT \NQPMV$).

\nop{+++++++++++++++++++++++++++++++++++++++++++++++++++++
\begin{fact}\label{prop-expr-DAm}
\begin{enumerate}
\item
$search(\DAm)= \NQPMV$,
\item
$opt(\DAm)= \OPT \NQPMV$.
\end{enumerate}
\end{fact}
\textbf{Proof} In \cite{Sac94b} it has been shown that a database
query $Q$ is defined by a query in $search(\DAm)$ if, and only if,
for each input database, the answers of $Q$ are
$\NP$-recognizable. Hence $search(\DAm)$ $= \NQPMV$ and
$opt(\DAm)= \OPT \NQPMV$.~\hfill $\Box$

\vspace{3mm}
The previous fact, firstly proved in \cite{GreSac97},
states that $\DAm$ captures the classes of $\NP$ search and
optimization queries.
++++++++++++++++++++++++++++++++++++++++++++++++++++++++++++++++++++}

\section{\DAsdc}

\noindent The problem in using \DAm\ to express search and
optimization problems is that the use of unrestricted negation
is often neither simple nor intuitive and, besides, it
does not allow expressive power (and complexity) to be controlled
and in some cases might also lead writing queries having no stable models.
In order to avoid these problems, we present a language,
called  \emph{\DAsdc}, where unstratified negation is
embedded into built-in constructs, so that the user is
forced to write programs using restricted forms of negation
without loss of expressive power.
Specifically, \emph{\DAsdc} extends \DAs\ with two simple
built-in constructs: head (exclusive) disjunction
and constraints, denoted by $\oplus$ and $\Leftarrow$,
respectively.

\paragraph{\bf Syntax.}

\noindent
In the following rules, $Body(\XX)$ and $Body(\XX,\YY,L)$ are conjunctions
of literals, whereas $\XX$ and $\YY$ are vectors of range restricted variables.

\vspace{2mm}
\noindent
An \emph{(exclusive) disjunctive rule} is
of the form:
\begin{eqnarray}\label{disjunctive-rule}
p_1(\XX_1) \oplus \cdots \oplus p_k(\XX_k) \leftarrow Body(\XX)
\end{eqnarray}

\noindent
where $\XX_i \subseteq \XX$ for all $i \in [1\.\.k]$.
The intuitive meaning of such a rule is that if $Body(\XX)$ is true,
then exactly one head atom $p_i(\XX_i)$ must be true.

\vspace{2mm} \noindent
A special form of disjunctive rule, called
\emph{generalized disjunctive rule}, of the form: \vspace{-1mm}
\begin{eqnarray}\label{generalized-disjunctive-rule}
\oplus_L \ p(\XX,L) \leftarrow Body(\XX,\YY,L)
\end{eqnarray}

\noindent is also allowed.
In this rule the number of head disjunctive atoms is not fixed,
but depends on the database instance and on the current computation (stable model).
The intuitive meaning of this rule is that the relation defined by
$\pi_{\bf X} Body(\XX,\YY,L)$ (the projection of the relation
$Body(\XX,\YY,L)$ on the attributes defined by $\XX$) is
partitioned into a number of subsets equal to the cardinality of
the  relation $\pi_{L} Body(\XX,\YY,L)$ (the number of distinct
values for the variable $L$). Some examples of generalized
disjunctive rules will be presented in the next section.

\vspace{2mm} \noindent A \emph{constraint (rule)} is of the form:
\begin{eqnarray}\label{constraint-rule}
\Leftarrow Body(\XX)
\end{eqnarray}

\noindent
A ground constraint rule is satisfied w.r.t. an interpretation $I$ if the
body of the rule is false in $I$. We shall often write constraints
using rules of the form $A_1 \vee \dots \vee A_k \Leftarrow
B_1,\dots,B_m$ (or  $B_1,\dots,B_m \Rightarrow A_1 \vee \dots \vee
A_k$) to denote a constraint of the form $\Leftarrow
B_1,\dots,B_m, \neg A_1,\dots,\neg A_k$ (i.e. negative literals
are moved from the body to the head). For instance, the constraint
$\tt \Leftarrow edge(X,Y), \neg v(X), \neg v(Y)$ of Example
\ref{Vertex-Cover} can be rewritten as $\tt v(X) \vee v(Y)
\Leftarrow edge(X,Y)$ or as $\tt edge(X,Y) \Rightarrow v(X) \vee
v(Y)$. Here the symbol $\vee$ denotes inclusive disjunction and is
different from $\oplus$, as the latter denotes exclusive
disjunction. It should be recalled that inclusive disjunction
allows more than one atom to be true while exclusive disjunction
allows only one atom to be true.

\begin{definition}\label{def-search-query1}
A \DAsdc\ {\em search query} is a pair $Q=\<\PP,g(t)\>$, where
$\PP$ is a \DAsdc\ program and $g(t)$ is an IDB atom.
A \DAsdc\ {\em optimization query} is a pair $\<\PP,$ $opt|g(t)|\>$,
where $opt$ is either $max$ or $min$. ~\punto
\end{definition}

\vspace*{2mm}
The query $\<\PP_2,{\tt v(X)}\>$ of Example \ref{Vertex-Cover} is a
\DAsdc\ search query, whereas the query $\<\PP_2,{\tt min|v(X)|}\>$ is a
\DAsdc\ optimization query.

\paragraph{\bf Semantics.}

\noindent The declarative semantics of a \DAsdc\ query
is given in terms of an `equivalent' \DAm\ query
and stable model semantics. Specifically, given a \DAsdc\ program
$\PP$, $st(\PP)$ denotes the standard \DAm\ program derived from
$\PP$ as follows:

\begin{enumerate}
\item
Every standard rule in $\PP$ belongs to $st(\PP)$,
\item
Every disjunctive rule $r \in \PP$ of the form
(\ref{disjunctive-rule}) is translated into $k$ rules of the form:

\[
\begin{array}{ll}
\hspace*{-5mm} p_j(\XX_j)\!\leftarrow & \!Body(\XX),\!\neg
p_{1}(\XX_{1}),\dots,\!\neg p_{j\mbox{-}1}(\XX_{j\mbox{-}1}),
\!\neg p_{j\mbox{+}1}(\XX_{j\mbox{+}1}),\dots,\!\neg p_k(\XX_k)
\end{array}
\]
with $j \in [1\.\.k]$, plus $((k-1) \times k)/2$ constraints of
the form:
\[
\begin{array}{ll}
\hspace*{-5mm} \Leftarrow & \!Body(\XX),p_i(\XX_i),p_j(\XX_j)
\end{array}
\]
with $i,j \in [1\.\.k]$ and $i < j$.
It is worth noting that the constraints are necessary only if $p_j$ is defined by some
other rule.
\item
Every generalized
disjunctive rule of the form (\ref{generalized-disjunctive-rule})
is translated into the two rules:

\[
\begin{array}{ll}
\hspace*{-7mm}
p(\XX,L)  \ \ \ \ \ \leftarrow & Body(\XX,\YY,L),\ \neg \mbox{\em diff}\_p(\XX,L) \\
\hspace*{-7mm} \mbox{\em diff}\_p(\XX,L) \leftarrow &
Body(\XX,\YY,L),\ p(\XX,L'),\ L' \neq L
\end{array}
\]
where {\em diff\_p} is a new predicate symbol and $L'$ is a new
variable, plus the constraint:

\[
\begin{array}{ll}
\hspace*{-5mm} \Leftarrow & Body(\XX,\YY,L),\ p(\XX,L_1), \
p(\XX,L_2), \ L_1 \neq L_2
\end{array}
\]

Here \emph{diff\_p} is used to avoid  inferring two ground atoms
$p(x,l_1)$ and $p(x,l_2)$ with $l_1 \neq l_2$.
Observe that even in this case the
constraint has to be introduced if $p$ is defined by some other
rule.
\item
Every constraint rule of the form (\ref{constraint-rule}) is
translated into a rule of the form:
\[
\hspace*{-5mm} c \leftarrow Body(\XX), \neg c
\]
\noindent where $c$ is a new predicate symbol not appearing
elsewhere.
\end{enumerate}

For any \DAsdc\ search query $Q = \<\PP,g(t)\>$ (resp. optimization
query $OQ = \<\PP,$ $opt|g(t)|\>$), $st(Q) = \<st(\PP),g(t)\>$ (resp.
$st(OQ) = \<st(\PP),opt|g(t)|\>$) \ denotes \ the \ corresponding \
\DAm\ (resp. optimization) query.

\begin{definition}
Given a \DAsdc\ query $Q$ and a database $D$, the
(non-deterministic) answer to the query $Q$ over $D$ is obtained by
applying the \DAm\ query $st(Q)$ to $D$, i.e. $Q(D) = st(Q)(D)$.
\hfill $\Box$
\end{definition}


%

It is worth noting that \DAsdc\ has the same
expressive power of $\DAm$, that is both $\NP$ search and optimization problems can be expressed
by means of \DAsdc\ queries under stable model semantics \cite{Zumpano04}.
The further restricted
languages \DAsd\ (Datalog with stratified negation and exclusive disjunction)
and \DAdc\ (Datalog with exclusive disjunction and constraints) have the same expressive power.
A similar result has been presented in \cite{EastTru05}, where it has been shown that
positive Datalog with constraints and head (inclusive) disjunction, called \emph{PS logic},
has the same expressive power as $\DAm$.
Clearly, \DAdc\ is captured by PS logic, since exclusive disjunction can be
emulated by using inclusive disjunction and constraints.
It is interesting to observe that analogous results could be obtained for others ASP languages.
For instance, (positive) Datalog with cardinality constrains, as proposed in Smodels,
captures the expressive power of $\DAm$ since exclusive disjunction
and denial constraints can be emulated by means of cardinality constraints \cite{Nie-etal99}.

\section{$\NPDA$}

\noindent
We now present a simplified version of \DAsdc\ introducing further restrictions on disjunctive rules.
The basic idea consists in restricting \DAsdc\ to obtain, without loss of
expressive power, a language which can be executed more
efficiently or easily translated in other formalisms.

\noindent
In the following rules, $Body(\XX,\YY)$ denotes a conjunction of
literals where $\XX$ and $\YY$ are vectors of range restricted
variables.

\vspace{2mm}
\noindent
A \emph{partition rule} is a disjunctive rule of the form:
\begin{eqnarray}\label{partition-rule}
p_1(\XX) \oplus \cdots \oplus p_k(\XX) \leftarrow Body(\XX,\YY)
\end{eqnarray}
or of the form
\begin{eqnarray}\label{partition-rule2}
p_0(\XX,c_1) \oplus \cdots \oplus p_0(\XX,c_k) \leftarrow
Body(\XX,\YY)
\end{eqnarray}
where $p_0,p_1,\dots,p_k$ are distinct IDB predicates not defined elsewhere
in the program and $c_1,\dots,c_k$ are
distinct constants. The intuitive meaning of these rules is that the
projection of the relation defined by $Body(\XX,\YY)$ on $\XX$ is partitioned
non-deterministically into $k$ relations or $k$ distinct sets of
the same relation. Clearly, every rule of form
(\ref{partition-rule2}) can be rewritten into a rule of form
(\ref{partition-rule}) and vice~versa.

\noindent
A \emph{generalized partition rule} is a
(generalized) disjunctive rule of the form:
\begin{eqnarray}\label{generalized-partition-rule}
\oplus_L \ p(\XX,L) \leftarrow Body(\XX,\YY), d(L)
\end{eqnarray}
where $p$ is an IDB predicate not defined elsewhere
and $d$ is a database domain predicate specifying the domain of
the variable $L$. The intuitive meaning of such a rule is that the
projection of the relation defined by $Body(\XX,\YY)$ on $\XX$ is
partitioned into a number of subsets equal to the cardinality of
the relation $d$.

\noindent In the following, the existence of \emph{subset rules} is
also assumed, i.e. rules of the form
\begin{eqnarray}
s(\XX) \subseteq Body(\XX,\YY)
\end{eqnarray}
where $s$ is an IDB predicate not defined elsewhere in the program.
Observe that a subset rule of the form above
corresponds to the generalized partition rule with $\tt d= \{0,1\}$.
On the other hand, every generalized partition rule can
be rewritten into a subset rule and constraints.

In the previous rules, $Body(\XX,\YY)$ is a
conjunction of literals not depending on predicates defined by
partition or subset rules.
We recall that the subset rules used here are based on the proposal in
\cite{GreSac97}. A similar type of subset rules has also been proposed
by \cite{Gel01} where the language ASET-Prolog (an extension of A-Prolog with sets)
is presented.


\begin{definition}
An \emph{\NPDA\ program} consists of three distinct sets of rules:
\begin{enumerate}
\item \emph{partition} and \emph{subset} rules defining \emph{guess (IDB) predicates},
\item standard \emph{stratified datalog rules} defining \emph{standard (IDB) predicates}, and
\item \emph{constraints rules}.
\end{enumerate}
where every guess predicate is defined by a unique subset or
partition rule. \hfill $\Box$
\end{definition}

According to the definition above, the set of IDB predicates of an \NPDA\
program can be partitioned into two distinct subsets (namely, guess and standard) depending
on the rules used to define them.
Clearly, predicates defined by partition or subset rules are
not recursive as the body of these rules cannot contain guess
predicates or predicates depending on guess predicates.

\begin{example}\label{vertex-cover-example3}\em
Vertex cover \em (version 3). The \NPDA\ program:
\[
\begin{array}{l}
\tt v(X) \subseteq node(X)\.  \\
\tt edge(X,Y) \Rightarrow  v(X) \vee v(Y)\.
\end{array}
\]
is derived from the one presented in Example \ref{Vertex-Cover} by
replacing the disjunctive rule defining $\tt v$ with a subset
rule. \hfill $\Box$
\end{example}

It is important to note that here a simpler form of \DAsdc\
queries is considered. Therefore, $\NPDA \subseteq \DAsdc$ and
every \NPDA\ query can be rewritten into an equivalent \DAm\
query.

\begin{definition}
An \NPDA\ \emph{search query} is a
pair $Q = \<\PP,g(t)\>$, where $\PP$ is an \NPDA\ program and
$g(t)$ is an IDB atom denoting the output relation.
An \NPDA\ \emph{optimization query} is a pair $\<\PP, opt|g(t)|\>$,
where $opt \in \{ max, min \}$.\punto
\end{definition}

Observe that, for the sake of simplicity, our attention is restricted to
optimization queries computing the maximum or minimum cardinality
of the output relation, although any polynomial function might be used.
Moreover, as stated by the following theorem, $\NPDA$ captures the
complexity classes of \NP\ search and optimization problems.

\begin{theorem}\label{Theorem-Expressiveness}\
\vspace*{-6mm}
\begin{enumerate}
\item
$search(\NP\ Datalog) = \NQPMV$, and 
\item
$opt(\NP\ Datalog)= \OPT \NQPMV$. 
\end{enumerate}
\end{theorem}
\textbf{Proof.}
Membership is trivial as $\NPDA\ \subseteq \DAsdc$.

\noindent
To prove hardness  the well-known Fagin's result is used
\cite{Fag74} (see also \cite{Joh90,Pap94}): it states that every
\NP\ recognizable database collection is defined by an existential
second order formula $\exists R \Phi$, where $R$ is a list of new
predicate symbols and $\Phi$ is a first-order formula involving
predicate symbols in a database schema $\DS$ and in $R$. As shown
in \cite{KolPap91}, this formula is equivalent to one of the form
({\em second order  Skolem normal form})

\begin{center}
$(\exists \SS)(\forall \XX)(\exists \YY) (\theta_1(\XX,\YY) \vee
\dots \vee \theta_k(\XX,\YY))$
\end{center}
where $\SS$ is a superlist of $R$,
$\theta_1,\dots,\theta_k$ are conjunctions of literals involving
variables in $\XX$ and $\YY$, and predicate symbols in $\SS$ and
$\DB$. Consider the program $\PP$:

\begin{center}$
\begin{array}{llll}
s_j(W_j) \oplus  \hat{s}_j(W_j) & \leftarrow &   & (\forall s_j \in \SS) \vspace{0.1cm} \\
q(X)             & \leftarrow & \theta_i(X,Y)   \hspace*{10mm} & (1 \leq i \leq k)\\
g                & \leftarrow & \neg q(X) &
\end{array}
$\end{center}

\noindent The first group of rules selects a set of constants from
the database domain, for each predicate symbol $s_j$. The second
group of rules implements the above second order formula. The
third rule checks if there is some $\XX$ for which the formula is
not satisfied. Therefore, the formula is satisfied if, and only
if, there is a stable model $M$ such that $\neg g \in M$.
\nop{++++++++++++++++++++++++++++++++++++++++++++++++++++++++++++
Completeness can be proved following the proof of Proposition
\ref{theo-expr-DAsdc} as the program used to simulate existential
second order formulae is a \NPDA\ program after replacing every
binary partition rule $s(W_1,\dots,W_k) \oplus
\hat{s}(W_1,\dots,W_k) \leftarrow$ with an equivalent subset rule
$s(W_1,\dots,W_k) \subset dom(W_1), \dots, dom(W_k)$ where
$dom(W_i)$ is the domain associated with the variable $W_i$.
++++++++++++++++++++++++++++++++++++++++++++++++++++++++++++++}
\hfill $\Box$

\vspace*{2mm} Thus, \NPDA\ has the same expressive power as both
\DAsdc\ and \DAm. The idea underlying \NPDA\ is that \NP\ search
and optimization problems can be expressed using partition (or
subset) rules to guess partitions or subsets of sets, whereas
constraints are used to verify properties to be satisfied by
guessed sets or sets computed by means of stratified rules.
It is important to observe that the proof of Theorem \ref{Theorem-Expressiveness} follows
a schema which has been used in other proofs concerning the expressive power
of Datalag with negation under stable model semantics \cite{Schlipf95,Sac94b,Baral03}.
Indeed, in such proofs (see for instance the proof of Theorem 6.3.1
in \cite{Baral03}) negation is only used to
express exclusive disjunction and constraints.

\vspace*{2mm} Although the aim of this work is not the definition
of techniques for the efficient computation of queries, we would
point out that \NPDA\ programs can be computed following the
classical stratified fixpoint algorithm enriched with a guess and
check technique.

\vspace*{2mm} The advantage of expressing  search and optimization
problems by using rules with built-in predicates rather than
standard \DAm\ rules is that the use of built-in atoms preserves
simplicity and intuition in expressing problems and allows queries
to be easily optimized and translated into other target languages
for which efficient executors exist.
%
A further advantage is that the use of built-in predicates in
expressing optimization queries permits us to easily identify
problems for which ``approximate" answers can be found in
polynomial time. For instance, maximization problems defined by
constraint free \NPDA\ queries where negation is only applied to
guess atoms or atoms not depending on guess atoms (called
deterministic) are constant approximable \cite{GreSac97}.
Indeed, these problems belong to the class of constant
approximable optimization problems $MAX \, \Sigma_1$\footnote{This
class was firstly introduced in \cite{PapYan91} as $MAX\ \NP$.}
\cite{KolTha95} and, therefore, \NPDA\ could also be used to define the
class of approximable optimization problems, but this is
outside the scope of this paper.

\NPDA\ allows us to also use a finite subset of the integer domain
and the standard built-in arithmetic operators.
More specifically, reasoning and computing over a finite set of integer
ranges is possible with the unary predicate \textbf{integer}, which consists
of the facts $\tt integer(x)$, with $\tt MinInt \leq x \leq MaxInt$,
and the standard arithmetic operators defined over the integer domain.

The following example shows how the arithmetics operators could be used to
compute prime numbers

\[
\begin{array}{l}
\tt composite(X) \leftarrow integer(Y), \ integer(Z), \ X = Y*Z\. \\
\tt prime(X) \leftarrow integer(X),\ not\ composite(X)\.
\end{array}
\]

Thus, the language allows arithmetic expressions which involve variables taking
integer values to appear as operands of comparison operators (see Example~\ref{ex:n-queens}).


\subsection*{Examples}

\noindent
Some examples are now presented showing how classic
search and optimization problems can be defined in \NPDA.

\begin{example}\label{Max-Satisfiability}
\emph{Max satisfiability.} Two unary relations $\tt c$
and $\tt a$ are given in such a way that a fact $\tt c(x)$ denotes that $\tt x$ is a
clause and a fact $\tt a(v)$ asserts that $\tt v$ is a variable
occurring in some clause. We also have two binary relations $\tt
p$ and $\tt n$ such that the facts $\tt p(x,v)$ and $\tt n(x,v)$
state that a variable $\tt v$ occurs in the clause $\tt x$
positively or negatively, respectively. A boolean formula, in
conjunctive normal form, can be represented by means of the
relations $\tt c$, $\tt a$, $\tt p$ and $\tt n$.

\noindent
The maximum number of clauses simultaneously satisfiable under
some truth assignment can be expressed by the query $\tt
\<\PP_{sat},max|f(X)|\>$ where $\tt \PP_{sat}$ is the following
program:
\[
\begin{array}{lll}
\tt s(X) & \subseteq & \tt a(X)\. \\
\tt f(X) & \leftarrow & \tt p(X,V), \ s(V)\. \\
\tt f(X) & \leftarrow & \tt n(X,V), \ \neg s(V)\.
\mbox{\hspace{73mm}} \Box
\end{array}
\]
\end{example}

Observe that the max satisfiability problem is constant
approximable as no constraints are used and negation is
applied to guess atoms only.

\vspace{2mm}
 In the following examples,  a database
graph $G=\<N,E\>$ defined by means of the unary relation $\tt
node$ and the binary relation $\tt
edge$ is assumed.

\begin{example}\label{3-Coloring}
\emph{k-Coloring.} Consider the well-known
problem of  k-colorability consisting in finding a
\emph{k-coloring}, i.e. an assignment of one of $k$ possible
colors to each node of a graph $G$
 such that no two adjacent nodes
have the same color. The problem can be expressed by means of the
\NPDA\ query $\<\tt \PP_{k\-col},col(X,C)\>$ where $\tt
\PP_{k\-col}$ consists of the following rules:

\[
\begin{array}{l}
\tt \oplus_C\ col(X,C) \leftarrow node(X), color(C)\. \\
\tt \Leftarrow edge(X,Y),\ col(X,C),\ col(Y,C)\.
\end{array}
\]
\noindent
and the base relation $\tt color$ contains exactly $k$ colors.
The first rule guesses an assignment of colors to the nodes of the graph,
while the constraint verifies that two joined vertices do not have the
same color.~\hfill $\Box$
\end{example}

\begin{example}\label{Min-Coloring}
\emph{Min Coloring.} The  query modeling the Min Coloring
problem is obtained from the the k-coloring example by
 adding a rule storing the
used colors as follows:
\[
\begin{array}{l}
\tt \oplus_C \ col(X,C)  \leftarrow  node(X), color(C)\. \\
\tt \Leftarrow edge(X,Y),\ col(X,C),\ col(Y,C)\. \\
\tt used\_color(C)  \leftarrow col(X,C)\.
\end{array}
\]

\noindent and replacing the query goal with  $\tt
min|used\_color(C)|$. \hfill $\Box$
\end{example}

\begin{example}\label{Min-Dominating-Set}
\emph{Min Dominating Set.} Given a graph $G=\<N,E\>$,  a subset of
the vertex set $V \subseteq N$ is a dominating set if for all $u
\in N-V$  there is a $v \in V$ such that $(u,v) \in E$. The \NPDA\
query $\<\tt \PP_{ds},v(X)\>$ expresses the problem of finding a dominating set,
where $\tt\PP_{ds}$ is the following program:
\[
\begin{array}{l}
\tt v(X)               \tt \subseteq   \tt node(X)\. \\
\tt connected(X)       \tt \leftarrow  \tt edge(X,Y), \ v(Y)\. \\
\tt node(X) \wedge \neg v(X)           \tt \Rightarrow \tt connected(X)\.
\end{array}
\]

\noindent The constraint states that every node not belonging to
the dominating set, namely the relation $\tt v$, must be
connected to some node in $\tt v$. A dominating set is said to be
$minimum$ if its cardinality is minimum. Therefore, the
optimization problem is expressed by replacing the query goal $\tt
v(X)$  with $\tt min |v(X)|$. \hfill $\Box$
\end{example}

Note that if an \NP-minimization query has an empty answer there
is no solution for the associated search problem.

\begin{example}\label{Min-Edge-Dominating-Set}
\emph{Min Edge Dominating Set.} Given a graph $G=\<N,E\>$,  a
subset of the edge set $A \subseteq E$ is an edge dominating set
if for all $e_1 \in E-A$  there is an $e_2 \in A$ such that $e_1$
and $e_2$ are adjacent. The min edge dominating set problem is
defined by the \NPDA\ query $\<\tt \PP_{eds}, min |e(X,Y)|\>$
where $\tt \PP_{eds}$ consists of the following rules:
\[
\begin{array}{l}
\tt e(X,Y) \subseteq edge(X,Y)\. \\
\tt v(X) \leftarrow e(X,Y)\. \\
\tt v(Y) \leftarrow e(X,Y)\.  \\
\tt edge(X,Y) \Rightarrow  v(X) \vee v(Y)\.
\end{array}
\]
\vspace*{-5mm} \hfill $\Box$
\end{example}

\begin{example}\label{ex:n-queens}
\emph{N-Queens.}
This problem consists in placing $N$ queens on an $N \times N$
chessboard in such a way that no two queens are in the same row, column, or diagonal.
It can be expressed by the \NPDA\ query $\<\tt \PP_{queen}, queen(R,C)\>$
where $\tt \PP_{queen}$ consists of the following rules:
\[
\begin{array}{l}
\tt \oplus_C\ queen(R,C) \leftarrow num(R), num(C)\. \\
\tt \Leftarrow queen(R_1,C), queen(R_2,C), R_1 \neq R_2.\\
\tt \Leftarrow queen(R_1,C_1), queen(R_2,C_2), R_1 \neq R_2, R_1+C_1=R_2+C_2.\\
\tt \Leftarrow queen(R_1,C_1), queen(R_2,C_2), R_1 \neq R_2, R_1-C_1=R_2-C_2.\\
\end{array}
\]
The database contains facts of the form $\tt num(1) \dots num(N)$ for the $\tt N$-queens problem.
The partition rule assigns to each row exactly one queen.
The first constraint states that no two different queens are in the same column.
The last two constraints state that no two different queens are on the same diagonal.~\hfill$\Box$
\end{example}

\begin{example}\label{ex:latin-squares}
\emph{Latin Squares.}
This problem consists in filling an $N \times N$ table with $N$ different symbols in such a way
that each symbol occurs exactly once in each row and exactly once in each column.
Tables are partially filled.
The \NPDA\ query $\<\tt \PP_{ls}, square(R,C,V)\>$ expresses the problem,
where $\tt \PP_{ls}$ consists of the following rules:
\[
\begin{array}{l}
\tt \oplus_V\ square(R,C,V) \leftarrow num(R), num(C), num(V) \\
\tt \Leftarrow square(R,C_1,V), square(R,C_2,V), C_1\neq C_2 \\
\tt \Leftarrow square(R_1,C,V), square(R_2,C,V), R_1\neq R_2 \\
\tt square(R,C,V) \Leftarrow preassigned(R,C,V) \\
\end{array}
\]
The database contains facts of the form $\tt num(1) \dots num(N)$
for an $\tt N \times N$ table and facts of the form
$\tt preassigned(R,C,V)$ whose meaning is that the entry $\tt \<R,C\>$ of the table contains
the symbol $V$ (here the symbols used are the numbers from 1 to $\tt N$).
The partition rule assigns exactly one symbol to each entry of the table.
The first (resp. second) constraint states that a symbol cannot occur more than once in the same row (resp. column).
The last constraint states that preassigned symbols must be respected.~\hfill$\Box$
\end{example}

\section{Translating ${\cal NP\ D}atalog$ Queries into OPL Programs}

Several languages have been designed and implemented for hard search and optimization problems.
These include logic languages based
on stable models (e.g. $DeRes$, \emph{DLV}, $ASSAT$, \emph{Smodels},
\emph{Cmodels}, \emph{Clasp})
\cite{DeRes,dlv,ASSAT,smodels,cmodels1,cmodels2,clasp},
constraint logic programming systems (e.g. SICStus Prolog, ECLiPSe, XSB, Mozart)
\cite{sicstusSys,WalSch*99,Warren-XSB,VanRoy99} and constraint programming
languages (e.g. ILOG OPL, Lingo) \cite{VAN99,Finkel*04}. The
advantage of using  logic languages based on stable model
semantics with respect to constraint programming is their ability
to express complex $\NP$ problems in a declarative way. On the
other hand, constraint programming languages are very efficient in
solving optimization problems.

As \NPDA\ is a language to express $\NP$ problems, the
implementation of the language can be performed by translating
queries into target languages specialized in combinatorial optimization problems,
such as constraint programming languages.
The implementation of \NPDA\ is carried out by means of a
system prototype translating \NPDA\ queries into OPL programs. OPL
is a constraint programming language well-suited for solving both search and
optimization problems. OPL programs are computed by means of the
ILOG OPL Development Studio \cite{Ilog}. This section shows how \NPDA\
queries are translated into OPL programs.

\NPDA\ programs have an associated database schema specifying the
used database domains and for each base predicate the  domain
associated with each attribute. For instance, the database schema
associated with the min coloring query of Example~\ref{Min-Coloring} is:
\[
\begin{array}{ll}
\tt DOMAINS:    & \tt node; color\. \\
\tt PREDICATES: & \tt edge(node,node)\. 
\end{array}
\]
\noindent
Starting from the database schema, the compiler also deduces the
schema of every derived predicate and introduces new domains,
obtained from the database domains. For instance, for the program
of Example~\ref{Min-Coloring} the schemas associated with the
predicates $\tt col$ and $\tt used\_color$ are, respectively, $\tt
col(node,color)$ and $\tt used\_color(color)$. Considering the
program of Example~\ref{Min-Coloring} and assuming to also have
the following rules:
\[
\begin{array}{l}
\tt p(X) \leftarrow node(X)\. \\
\tt p(X) \leftarrow color(X)\. \\
\tt q(X) \leftarrow node(X), color(X)\.
\end{array}
\]
the schemas associated with $\tt p$ and $\tt q$ are $\tt p(D_p)$
and $\tt q(D_q)$ where $\tt D_p$ is the union of the domains $\tt node$ and $\tt color$,
whereas $\tt D_q$ is the intersection of the domains $\tt node$ and $\tt color$.
Database domain instances are defined by means of unary ground facts.
Integer domains are declared differently.
For instance, the database schema associated with the \emph{N-queens} program of Example~\ref{ex:n-queens} is
as follows:
\[
\begin{array}{ll}
\tt INT\-DOMAINS:    & \tt num\, \. \\
\end{array}
\]

Moreover, whenever the \textbf{integer} predicate is used in a program, the range of considered integers has to be
specified in the schema, as shown in the following example:
\[
\begin{array}{ll}
\tt MinInt=0\.\\
\tt MaxInt=10\. \\
\end{array}
\]

A predicate $p$ is said to be \emph{constrained} if
i) $p$ depends on a guess predicate, and
ii) there is a constraint or an optimized query goal containing $p$ or containing a
predicate $q$ which depends on $p$.
Moreover, a constrained predicate is said to be \emph{recursion-dependent} if it is recursive or
depends on a constrained recursive predicate.

Every \NPDA\ program $\PP$ consists of
a set $\PP_S$ of standard rules,
a set $\PP_G$ of rules defining guess predicates and
a set $\PP_C$ of constraints.
A program $\PP = \PP_S \cup \PP_G \cup \PP_C$ can be also partitioned into four sets:
\begin{enumerate}
\item
$\PP^1 = \PP^1_S$ consisting of the set of rules
defining standard predicates not depending on guess predicates;
\item
$\PP^2 = \PP_G \cup \PP^2_S \cup  \PP^2_C$ consisting of
(i) the set of  rules defining guess predicates ($\PP_G$),
(ii) the set of standard rules defining constrained predicates which are not recursion-dependent
($\PP^2_S$) and
(iii) the set of constraints $\PP^2_C$ containing only
base predicates and predicates defined in $\PP_G \cup \PP^1_S \cup \PP^2_S $;
\item
$\PP^3 = \PP^3_S \cup  \PP^3_C$ consisting of
the set of standard rules defining constrained, recursion-dependent predicates
($\PP^3_S$) and the set of constraints
$\PP^3_C$ containing predicates defined in $\PP^3_S $;
\item
$\PP^4 = \PP^4_S$ consisting of the set of rules defining standard
predicates which depend on guess predicates and are not constrained.
\end{enumerate}
%

The evaluation of an \NPDA\ program $\PP$ over a database $\D$
is carried out by performing the following steps:
\begin{enumerate}
\item
Firstly, the (unique) stable model of $\<\PP^1_S, \D\>$
(say it $\D \cup M_1$) is computed.
\item
Next, a stable model of $\<\PP^2_S
\cup \PP_G, \D \cup M_1 \>$ satisfying the constraints $\PP^2_C$
(say it $\D \cup M_1 \cup M_2$) is computed.
\item
Afterwards, if a model $\D \cup M_1 \cup M_2$ exists,
the (unique) stable model of
$\<\PP^3_S, \D \cup M_1 \cup M_2 \>$
(say it $\D \cup M_1 \cup M_2 \cup M_3$) is computed.
If this model satisfies the constraints $\PP^3_C$, then the next step is executed,
otherwise the second step is executed again, that is, another
stable model of $\<\PP^2_S \cup \PP_G, \D \cup M_1 \>$ satisfying the constraints $\PP^2_C$
is computed.
\item
Finally, if a model $\D \cup M_1 \cup M_2 \cup M_3$ satisfying the constraints $\PP^3_C$
exists, the (unique) stable model of $\<\PP^4_S, \D \cup M_1 \cup M_2 \cup M_3\>$ is evaluated.
\end{enumerate}

\noindent
It is worth noting that, if there is no constrained recursive predicate,
the component $\PP^3$ is empty and then an \NPDA\ program can be evaluated
by performing only steps 1,2 and 4
(that is, the iteration introduced in step~3 is not needed).

The partition of programs into four components suggests that
subprograms $\PP^1_S$, $\PP^3_S$ and $\PP^4_S$ can be evaluated by means of the
standard fixpoint algorithm. In the following, stratified
subprograms, such as $\PP^1_S$, $\PP^3_S$ and $\PP^4_S$, are called
\emph{deterministic} as they have a unique stable model, whereas
subprograms which may have zero or more stable models are called
\emph{non-deterministic}.
Thus, given a database $\D$ and an \NPDA\ query $Q=\<P,G\>$, we have
to generate an OPL program equivalent to the application of the
query $Q$ to the database $\D$.

\noindent  We first show how the database is
translated and next consider the translation of queries.

\paragraph{Database translation.}

An integer domain relation is translated into a set of integers,
whereas a non-integer domain relation is translated into a set of strings.
The translation of a base relation with arity $n>0$ consists of
two steps:
(i) declaring a new tuple type with $n$ fields (whose type is either string or integer, according to the schema),
(ii) declaring a set of tuples of this type.
For instance, the translation of the database containing the facts
$\tt node(a), node(b),$ $\tt node(c),$ $\tt node(d),$  $\tt
edge(a,b),$ $\tt edge(a,c),$ $\tt edge(b,c)$ and $\tt edge(c,d)$,
consists of the following OPL declarations:

\begin{quote}

$\tt \{ {\bf string} \}\ node \ = \{ a, b, c, d \}$;

$\tt {\bf tuple}\ edge\_type \ \{ {\bf string}\ a_1;\ {\bf string}\ a_2; \};$

$\tt \{ edge\_type\}\ edge = \{ \<a,b\>, \<a,c\>,
\<b,c\>, \<c,d\> \};$
\end{quote}

The database $\tt num(1), num(2), num(3)$ for the \emph{N-Queens} problem of Example~\ref{ex:n-queens}
is translated as follows:

\begin{quote}
$\tt \{ {\bf int} \}\ num \ = \{ 1, 2, 3 \}$;
\end{quote}

When an integer range is specified in the schema, the following set is added to the OPL database:

\begin{quote}
$\tt \{\textbf{int}\}\ integer\ =\ asSet(MinInt\ \.\.\ MaxInt);$
\end{quote}

where $\tt MinInt$ and $\tt MaxInt$ are the values specified in the schema.

\paragraph{Query translation.}
The translation of an \NPDA\ query $Q = \<\PP,G\>$ is carried out
by translating the deterministic subprograms into \emph{ILOG OPL Script} programs
by means of a function $Fixp$ and the non-deterministic subprograms
into OPL programs by means of a function $\WP$ or a slightly
different function $\WQ$ if the predicate in the query goal is defined in $\PP^2$.
More specifically, $Fixp(\PP)$ generates an OPL script program
which emulates the fixpoint computation of $\PP$, whereas
$\WP(\PP)$ (resp. $\WQ(\QQ)$) translates the \NPDA\ program $\PP$
(resp. query $\QQ$) into an equivalent OPL program.

\vspace{2mm}
\noindent
It is worth noting that:
\begin{enumerate}
\item
If the query goal is not defined over component $\PP^4$, this
component does not need to be evaluated and, therefore, it is not
translated into an OPL Script program.
\item
If the query goal is defined in component $\PP^1$,
we have to check that the components $\PP^2$ and $\PP^3$ admit stable models.
\item
If the query goal $G$ is defined in component $\PP^2$,
we have to compute the query $\<\PP^2, G\>$ over the stable model (which includes the database)
obtained from the computation of component $\PP^1$ and
check that component $\PP^3$ admits stable models.
\item
Similarly, if the query goal $G$ is defined in component $\PP^3$,
we have to compute the query $\<\PP^3, G\>$ over a stable model of
$\PP^1 \cup \PP^2 \cup \DB$.
\item
If the query goal $G$ is defined in component
$\PP^4$, first we compute a stable model $M$ for components
$\PP^1$, $\PP^2$ and $\PP^3$ and next compute the fixpoint
of component $\PP^4$ over $M$.
\end{enumerate}

First, we informally present how a deterministic
component ($\PP^1_S$, $\PP^3_S$ and $\PP^4_S$ in our partition) is translated
into an \emph{ILOG OPL Script} program, and next we show how the remaining rules
are translated into an OPL program.

%


\paragraph{Translation of deterministic components.}

The translation of a stratified program $\PP_S$ produces an
ILOG OPL Script program which emulates the application of the naive
fixpoint algorithm to the rules in $\PP_S$.

\noindent
The following example shows how a set of stratified rules
is translated into an ILOG OPL Script program.

\begin{example}\label{transitive-closure-example}
\emph{Transitive closure.}
Consider the following $\NPDA$ program $\PP_{tc}$
computing the transitive closure of a graph:
\[
\begin{array}{l}
\tt tc(X,Y) \leftarrow edge(X,Y). \\
\tt tc(X,Y) \leftarrow edge(X,Z), tc(Z,Y).
\end{array}
\]

\noindent
The corresponding OPL Script program $Fixp(\PP_{tc})$ is as follows:

\[
\small
\begin{array}{l}
\hspace*{0mm}\it //\ tc\ declaration \\
\hspace*{0mm}\tt \textbf{int}\ tc\,[node][node]; \vspace*{2mm} \\
\hspace*{00mm}\tt    \textbf{execute} \{ \\
\hspace*{05mm}\it    //\ exit\ rule \\
\hspace*{05mm}\tt    \textbf{for}\ (\textbf{var}\ x\ \textbf{in}\ edge)\ \{ \\
\hspace*{10mm}\tt      tc\,[x\.a1][x\.a2]=1;\\
\hspace*{05mm}\tt    \} \vspace*{2mm} \\
\end{array}
\]
\vspace*{-5mm}
\[
\small
\begin{array}{l}
\hspace*{5mm}\it    //\ recursive\ rule \\
\hspace*{5mm}\tt    \textbf{var}\ modified = \textbf{true}; \\
\hspace*{5mm}\tt    \textbf{while}\ (modified)\ \{ \\
\hspace*{10mm}\tt       modified = \textbf{false}; \\
\hspace*{10mm}\tt       \textbf{for}\ (\textbf{var}\ e\ \textbf{in}\ edge) \\
\hspace*{15mm}\tt          \textbf{for}\ (\textbf{var}\ y\ \textbf{in}\ node) \\
\hspace*{20mm}\tt             \textbf{if}\ (tc\,[e\.a2][y] == 1\ \&\ tc\,[e\.a1][y]==0)\ \{\\
\hspace*{30mm}\tt                     tc\,[e\.a1][y] = 1; \\
\hspace*{30mm}\tt                     modified=\textbf{true}; \\
\hspace*{20mm}\tt                \} \\
\hspace*{5mm}\tt    \} \\
\hspace*{0mm}\tt    \} \\
\end{array}
\]
\vspace*{-5mm}
\hfill $\Box$
\end{example}

\noindent

In the program above we have three sets of statements declaring
variables and computing exit and recursive rules. Specifically:
\begin{enumerate}
\item
A two-dimensional integer array $\tt tc$ is declared.
\item
The first $\tt \bf forall$
block evaluates the exit rule defining $\tt tc$ by inserting each
edge into the transitive closure.
\item
The recursive rule is
evaluated by means of the classical \emph{naive fixpoint algorithm}~
\cite{Ull88}. Specifically, the statements inside the $\tt \bf
while$ block insert a pair $\<e\.a1,y\>$ in the transitive closure, if
there exist an edge $\<e\.a1,e\.a2\>$ and a node $y$ such that the
transitive closure contains the pair $\<e\.a2,y\>$. The loop ends when
no more pairs of nodes can be derived.
\end{enumerate}

If a program contains negated literals, it is possible to apply
the stratified fixpoint algorithm, by dividing the rules into
strata and computing one stratum at a time, following the order
derived from the dependencies among predicate symbols.
%

\paragraph{Translation of non-deterministic components.}

The translation of a non-deterministic program $\PP$ (denoted by
$\WP(\PP)$) produces an OPL program. For the sake of simplicity of
presentation, it is assumed that  $\PP$ satisfies the
following conditions:
\begin{itemize}
\item
guess predicates are defined by either generalized partition
rules or subset rules;
\item
standard predicates are defined by a
unique extended rule of the form:
\[
\tt A \leftarrow body_1 \vee \cdots \vee body_m
\]
where $\tt body_i$ is a conjunction of literals;
\item
constraint rules are of the form $\tt A \Leftarrow B$, where $\tt A$ is a
disjunction of atoms and $\tt B$ is a conjunction of atoms;
\item
rules do not contain two (or more) occurrences of the same variable
taking values from different domains;
\item constants appear only in
built-in atoms of the form $\tt x\, \theta\, y$ where $\tt \theta$
is a comparison operator.
\end{itemize}

It should be noticed that the previous assumptions do not imply
any  limitation as every program can be
rewritten in such a way that it satisfies them.
For instance, the two rules defining the predicate $\tt f$ in
Example~\ref{Max-Satisfiability} can be rewritten into the rule
\[
\tt f(X) \leftarrow ( p(X,V), s(V)) \vee (n(X,Z), \neg s(Z))
\]
\noindent whereas the rules defining the predicate $\tt v$ in
Example \ref{Min-Edge-Dominating-Set} can be rewritten in the form
\[
\tt v(X) \leftarrow e(X,V) \vee e(U,X)
\]

Specifically, the function $\WP$ receives in input a program $\PP$
and gives in output an OPL program consisting of two components
$\WP(\PP) = ( \TD(\PP), \TP(\PP$)) where (i) $\TD(\PP)$ consists
of the definition of arrays of integers and decision variables, (ii) $\TP(\PP)$ translates the
\NPDA\ program into an OPL program. Analogously, the function $\WQ$
receives in input an \NPDA\ query $\QQ = \<\PP,G\>$ and gives in output an
OPL program consisting of two components $\WQ(\<\PP,G\>) = (
\TD(\PP), \TQ(\<\PP,G\>$)) where $\TQ(\<\PP,G\>)$ translates the
\NPDA\ query into an OPL program.

\noindent
The function $\TD(\PP)$ introduces some data structures for each IDB predicate defined in $\PP$.
Specifically, for each IDB predicate $\tt p$ defined in $\PP^2$ with arity $k$,
a $k$-dimensional array of boolean decision variables is introduced as
follows:

\begin{quote}
$\tt \textbf{dvar\ boolean}\ \ p[D_1,\dots,D_k];$
\end{quote}

\noindent
where $\tt D_1,\dots,D_k$ denote the domains on which the
predicate $\tt p$ is defined. For instance, for the binary
predicate $\tt col$ of Example \ref{Min-Coloring} the declaration

\begin{quote}
$\tt \textbf{dvar\ boolean}\ \ col[node, color];$
\end{quote}
\noindent
is introduced.
For any other IDB predicate $\tt q$ defined in $\PP$ with arity $m$,
a $m$-dimensional array of integers is introduced as
follows:
\begin{quote}
$\tt \textbf{int}\  q[D_1,\dots,D_m];$
\end{quote}
\noindent
where $\tt D_1,\dots,D_m$ denote the domains on which the
predicate $\tt p$ is defined.

\vspace*{3mm}
\noindent
The function $\TQ$ and $\TP$ are defined as follows:

\begin{enumerate}
\item
\textbf{Query}:
$\tt \TQ(\<\PP, G\> ) = \TQ(G) \ \TP(\PP)  $

\item \textbf{Goal}:
\begin{enumerate}
\item
$\tt \TQ( v(X_1,\dots,X_k) ) =\ \emptyset$
\item
$\tt \TQ( min|v(X_1,\dots,X_k)| ) = \\
\hspace*{8mm} \textbf{minimize}\  \textbf{sum} (X_1 \ \textbf{in} \ dom(X_1) ,\dots, X_k \ \textbf{in} \ dom(X_k)) \ v[X_1,\dots,X_k];$
\item
$\tt \TQ( max|v(X_1,...,X_k)| ) = \\
\hspace*{8mm} \textbf{maximize}\ \textbf{sum} (X_1\ \textbf{in} \ dom(X_1) ,\dots, X_k \ \textbf{in} \ dom(X_k)) \ v[X_1,\dots,X_k];$
\end{enumerate}

\item \textbf{Sequence of rules}:
$\TP( S_1 \ \dots \ S_n ) = \textbf{subject to}\{\ \TP(S_1) \ \dots\  \TP(S_n) \};$

\items{2} \textbf{Partition rules} of the form
\begin{quote}
$\tt \tt \oplus_L s(X_1,\ldots,X_k,L) \leftarrow body(X_1,\ldots,X_k, Y_1,\ldots, Y_n), d(L)$
\end{quote}
\noindent
are translated into the following OPL statement:

\begin{quote}$ \tt
\noindent
\hspace*{0cm}  \textbf{forall} (X_1 \ \textbf{in}\ \ dom(X_1), \dots, \ X_k \ \textbf{in}\ dom(X_k)) \ \\ 
\hspace*{8mm}      \TP(\exists(Y_1, \dots, Y_n)\ body(X_1,\dots,X_k, Y_1,\dots, Y_n)) > 0 \\
\hspace*{16mm}  \Rightarrow \textbf{sum}(L \ \textbf{in} \ d) \ s[X_1,\dots, X_k,L]~==~1 ;\\
\hspace*{0cm}  \textbf{forall} (X_1 \ \textbf{in}\ dom(X_1), \dots, \ X_k \ \textbf{in}\ dom(X_k),\ L\ \textbf{in}\  d) \\
\hspace*{8mm}      s[X_1,\dots, X_k,L] > 0 \Rightarrow \TP(\exists(Y_1, \dots, Y_n)\ body(X_1,\dots,X_k, Y_1,\dots, Y_n)) > 0   ;  \\
$
\end{quote}
\vspace*{-3mm}


\item \textbf{Subset rules} of the form
\begin{quote}
$\tt s(X_1,\dots,X_k) \subseteq body(X_1,\dots,X_k, Y_1,\dots, Y_n)$
\end{quote}
are translated as follows:

\begin{quote}$ \tt
\noindent
\hspace*{0cm}  \textbf{forall} (X_1 \ \textbf{in}\ dom(X_1), \dots, \ X_k \ \textbf{in}\ dom(X_k)) \ \\
\hspace*{8mm}      s[X_1,\dots, X_k] > 0 \Rightarrow \TP(\exists(Y_1,\dots, Y_n)\ body(X_1,\dots,X_k, Y_1,\dots, Y_n)) > 0; \\
$
\end{quote}
\vspace*{-3mm}

\item \textbf{Standard rules} of the form
\begin{quote}
$\tt p(X_1,\dots,X_k) \leftarrow
Body_1(X_1,\dots,X_k,Y^1_1,\dots,Y^1_{n_1}) \vee \cdots \vee Body_m(X_1,\dots,X_k,Y^m_1,\dots,Y^m_{n_m})$
\end{quote}
are translated as follows:

\begin{quote}$ \tt
\noindent
\hspace*{0cm}  \textbf{forall} (X_1\  \textbf{in} \ dom(X_1), \dots, X_k\  \textbf{in} \ dom(X_k)) \ \\
\hspace*{6mm}         p[X_1, \dots, X_k] > 0 \Leftrightarrow
                          \TP(\exists(Y^1_1,\dots,Y^1_{n_1})Body_1) \mbox{+} \cdots \mbox{+}
                          \TP(\exists(Y^m_1,\dots,Y^m_{n_m})Body_m)~>~0; \vs\\
$
\end{quote}
\vspace*{-3mm}

where $\tt Y^i_1,\dots,Y^i_{n_i}$ is the list of existentially
quantified variables in $\tt Body_i$.

\items{2}
\textbf{Conjunction of literals with existentially quantified variables}:
A conjunction of literals with $\tt n > 0$ existentially quantified
variables is translated as follows:
\[
\tt \TP(\exists(Y_1,\ldots,Y_{n}) Body) = (\textbf{sum}(Y_1\
\textbf{in}\ D_1, \ldots,Y_{n}\ \textbf{in}\ D_{n})\
(\TP(Body)))
\]

where $\tt D_j$ is the domain associated with the variable $\tt
Y_j$.

\items{2} \textbf{Conjunction of literals without existentially quantified variables}:
\vspace*{1mm}

$\tt \TP(A_1,\ldots,A_k) = \left\{
\begin{array}{ll}
\tt (\TP(A_1) * \cdots * \TP(A_k)) & \mbox{ if $\tt k > 0$} \vspace*{2mm} \\
\tt 1 & \mbox{ if $\tt k = 0$}
\end{array}
\right.$

\item \textbf{Literal}:

$\tt \TP(q(X_1,\ldots,X_k)) = \left\{
\begin{array}{ll}
\tt q[X_1,\ldots,X_k], & \mbox{ if } \tt q \small\mbox{ is a derived pred.} \\
\tt (\textbf{sum}(\<X_1,\ldots,X_k\> \ \textbf{in}\ q)\ 1 > 0), & \mbox{ if } \tt q \small\mbox{ is a base pred.} \\
\end{array}
\right.$

\vspace*{2mm} $\tt \TP(q(X)) = (\textbf{sum}(X \ \textbf{in}\ q)\
1 > 0)$  if $\tt q$ is a domain predicate,

\vspace*{2mm} $\tt \TP(E_1\, \theta\ E_2) = (E_1\, \theta\, E_2)$, where
$\theta$ is a comparison operator and $\tt E_1,E_2$ are either variables or constants or arithmetic expressions,

\vspace*{2mm} $\tt \TP(\neg A) = (1 - \TP(A))$;

\items{2}
\textbf{Constraints} of the form \ $\tt A_1 \vee \cdots \vee
A_m \Leftarrow body(X_1, ..., X_k)$ where $\tt body(X_1, ...,
X_k)$ is a conjunction of atoms are translated as follows:

$\TP(\ {\tt A_1 \vee \cdots \vee A_m \Leftarrow body(X_1, \ldots,
X_k)}\ ) = $
\[
\begin{array}{l}
\hspace*{0mm}\tt     \textbf{forall} (X_1\  \textbf{in} \ dom(X_1), \ldots , X_k\  \textbf{in} \ dom(X_k)) \  \\
\hspace*{8mm}\tt      \TP(body(X_1, \ldots , X_k) ) >0 \ \Rightarrow \tt (\TP(A_1) + \cdots + \TP(A_m)) > 0; \vs\\
\end{array}
\]

For $\tt m=0$ the above constraint becomes
$\TP\tt (body(X_1, \ldots ,X_k) ) >0 \ \Rightarrow \tt \textbf{false};$.

\end{enumerate}

Observe that the OPL code associated with the translation of a
rule  can be  simplified by means of trivial reductions. As
an example, an expression of the form: $\tt ((c > 0) >0)$ can be
simply replaced by $\tt (c > 0) $, whereas expressions of the form
$\tt 1 * 1$ are replaced by $\tt 1$.

The following theorem shows the correctness of our translation. As
we partition a program $\PP$ into four distinct components
$\PP^1$, $\PP^2$, $\PP^3$ and $\PP^4$, where the components $\PP^1$,
$\PP^3_S$ and $\PP^4$ are computed by means of a fixpoint algorithm, whereas the
components $\PP^2$ and $P^3_C$ are translated into OPL programs, we next show
the correctness of the translation of queries $Q = \<\PP,G\>$,
where $\PP = \PP^2$, i.e. we assume that components $\PP^1$,
$\PP^3$ and $\PP^4$ are empty.
Thus, such queries consist only of rules defining guess predicates, constraints and
standard rules defining constrained predicates which are not recursion-dependent.
Programs and queries of this form
will be called $\cal R$-\NPDA\ (restricted \NPDA).

\begin{theorem}
For every  $\cal R$-\NPDA\ query $Q$, $\WQ(Q) \equiv Q$.
\end{theorem}

\noindent
\textbf{Proof.}\ \ For each $\cal R$-\NPDA\ query $Q = \<\PP,G\>$,
where each standard predicate is defined by a unique extended
rule, the query $Q^r = \<\PP^r,G^r\>$ is derived as follows:

\begin{enumerate}
\item every generalized partition rule of the form:
\[
\begin{array}{l}
\tt \oplus_L s(X_1,\dots,X_k,L) \leftarrow body(X_1,\dots,X_k,$
$\tt Y_1,\dots, Y_n), d(L)
\end{array}
\]
is substituted by the constraints:
\[
\begin{array}{l}
\tt body(X_1,\dots,X_k, Y_1,\dots, Y_n) \Rightarrow s(X_1,\dots,X_k,L) \\
\tt s(X_1,\dots,X_k,L_1), s(X_1,\dots,X_k,L_2) \Rightarrow L_1 = L_2\\
\tt s(X_1,\dots,X_k,L) \Rightarrow body(X_1,\dots,X_k, Y_1,
\ldots, Y_n)
\end{array}
\]
\item each subset rule of the form:
\[
\begin{array}{l}
\tt s(X_1,\dots,X_k) \subseteq body(X_1,\dots,X_k, Y_1, \dots,
Y_n)
\end{array}
\]
is replaced by the constraint:
\[\tt
s(X_1,\dots,X_k) \Rightarrow body(X_1,\dots,X_k, Y_1, \dots, Y_n)
\]
\item every standard rule of the form:
\[
\tt A \leftarrow Body_1 \vee \cdots \vee Body_m
\]
is replaced by the constraint\footnote{ A shorthand for the two
constraints:
\[
\begin{array}{l}
\tt A \Rightarrow Body_1 \vee \cdots \vee Body_m \\
\tt A \Leftarrow  Body_1 \vee \cdots \vee Body_m
\end{array}
\]
}: 
\[\tt
A \Leftrightarrow Body_1 \vee \cdots \vee Body_m
\]
\item For each derived predicate $\tt p$ with schema $\tt p(dom_1,
\dots, dom_k)$ we introduce (i) a new predicate symbol $\tt p'$
with schema $\tt p'(dom_1, \dots, dom_k)$, and (ii) a rule of the
following form:
\begin{equation}\label{guess-variable}
\tt p(X_1, \dots, X_k) \oplus p'(X_1, \dots, X_k) \leftarrow
dom_1(X_1), \dots, dom_k(X_k)
\end{equation}
These rules are introduced to assign, non-deterministically, a
truth value to derived atoms.
\end{enumerate}

\vspace{2mm} Clearly, the queries $Q$ and $Q^r$ are equivalent as
the correct truth value of derived atoms is determined by the
constraints. It is worth pointing out that for each (partition,
subset and standard) rule $r$  a constraint of the form $Head(r)
\Rightarrow Body(r)$ was introduced to guarantee that models
contain only ``supported atoms", i.e. atoms derivable from $r$.

The program $\WQ(Q)$ is just a translation of $Q^r$ into OPL
statements where:
\begin{itemize}
\item rules of form (\ref{guess-variable}) do not need to be
translated into correspondent OPL statements as each derived
predicate $\tt p$, defined by such a rule, is translated into a
boolean $k$-dimensional array. \item the first two constraints,
derived from the rewriting of partition rules,  for ensuring
(i)~the assignment of each element in the body to some class $L$
and (ii)~the uniqueness of this assignment, are rewritten into a
unique OPL constraint. \hfill $\Box$
\end{itemize}

\begin{example}\label{OPL-Min-COloring}
\emph{Min-Coloring.} The OPL program corresponding to the
(simplified) translation of the min-coloring query of Example
\ref{Min-Coloring} is as follows:

\begin{quote}\small
$ $\\
\textbf{dvar\ boolean}\ \ {\tt col[node,color];} \\
\textbf{dvar\ boolean}\ \ {\tt used\_color[color];} \vspace*{2mm}\\
\textbf{minimize} \vspace*{1mm}\\
\hspace*{1cm}     \textbf{sum}($\tt c$ \textbf{in} $\tt color)\ used\_color[c];$ \vspace*{1mm}\\
\textbf{subject to} \{ \\
\hspace*{1cm}     \textbf{forall} ($\tt x$ \textbf{in} $\tt node$) \\
\hspace*{2cm}         $\tt (sum\ (x\ \textbf{in}\ node)\ 1>0) > 0$ \ $\Rightarrow \tt \textbf{sum}(c\ \textbf{in}\ color)\ col[x,c]==1$; \vspace*{2mm} \\
\hspace*{1cm}     \textbf{forall} ($\tt x$ \textbf{in} $\tt node$, $\tt c$ \textbf{in} $\tt color$) \\
\hspace*{2cm}         $\tt col[x,c] > 0 \Rightarrow (sum(x\ \textbf{in}\ node)\ 1>0) > 0$; \vspace*{2mm} \\
\hspace*{1cm}     \textbf{forall} ($\tt c$ \textbf{in} $\tt color$)  \\
\hspace*{2cm}        $\tt used\_color[c] > 0 \Leftrightarrow $ \textbf{sum}($\tt x$ \textbf{in} $\tt node$) $\tt col[x,c] > 0 ;$ 
\vspace*{2mm} \\
\hspace*{1cm}     \textbf{forall} ($\tt x$ \textbf{in} $\tt node,\ y$ \textbf{in} $\tt node,\ c$ \textbf{in} $\tt color$)  \\
\hspace*{2cm}         (\textbf{sum}($\tt \<x,y\>$ \textbf{in} $\tt edge)\ 1 > 0) * col[x,c] * col[y,c] > 0 \Rightarrow $ \textbf{false}; \\
\};
\hfill $\Box$
\end{quote}
\end{example}

\paragraph{Code optimization.}
The number of (ground) constraints can be strongly reduced by
applying simple optimizations to the OPL code.

\begin{itemize}

\item
\emph{Range restriction.}
If the OPL code contains constructs of the form:

\begin{quote}
\hspace*{1cm} $\tt\textbf{forall}(X_1\ \textbf{in}\ D_1,\dots, X_n\ \textbf{in}\ D_n)$ \\
\hspace*{2cm}      $(\tt\textbf{sum}(\<X_1,\dots,X_k\>\ \textbf{in}\ T)\ 1 > 0 )\ \<Statement_1\> \Rightarrow \<Statement_2\>$
\end{quote}

the \textbf{sum}
construct can be deleted so that the constraint can be rewritten
as follows:

\begin{quote}
\hspace*{1cm} $\tt\textbf{forall}(\<X_1,\dots,X_k\>\ \textbf{in}\ T,\ X_{k+1}\ \textbf{in}\ D_{k+1},\dots, X_n\ \textbf{in}\ D_n)$ \\
\hspace*{2cm}      $\tt 1\ \ \<Statement_1\> \Rightarrow \<Statement_2\>$
\end{quote}

\noindent
If the OPL code contains constructs of the form:
\begin{quote}
\hspace*{1cm} $\tt\textbf{forall}(X_1\ \textbf{in}\ D_1,\dots, X_n\ \textbf{in}\ D_n)$ \\
\hspace*{2cm}     $\tt(\textbf{sum}(X_1\ \textbf{in}\ D_1)\ 1 > 0 )\ \ \<Statement\>$
\end{quote}

the \textbf{sum} construct can be deleted so that the constraint can be rewritten as:
\begin{quote}
\hspace*{1cm} $\tt\textbf{forall}(X_1\ \textbf{in}\ D_1,\dots, X_n\ \textbf{in}\ D_n)$ \\
\hspace*{2cm}      $\tt 1\ \ \<Statement\>$
\end{quote}

\begin{example}\label{Min-COloring-Opt1}
By applying the optimizations above, the min-coloring problem can be
rewritten as follows:

\begin{quote}\small
$ $\\
\textbf{dvar\ boolean}\ {\tt\ col[node,color];} \\
\textbf{dvar\ boolean}\ {\tt\ used\_color[color];} \vspace*{2mm}\\
\textbf{minimize} \vspace*{1mm}\\
\hspace*{1cm}     \textbf{sum}($\tt c$ \textbf{in} $\tt color)\ used\_color[c];$ \vspace*{1mm}\\
\textbf{subject to} \{ \\
\hspace*{1cm}     \textbf{forall} ($\tt x$ \textbf{in} $\tt node$) \\
\hspace*{2cm}         $1 > 0$ \ $\Rightarrow \tt \textbf{sum}(c\ \textbf{in}\ color)\ col[x,c]==1$; \vspace*{2mm} \\
\hspace*{1cm}     \textbf{forall} ($\tt x$ \textbf{in} $\tt node$, $\tt c$ \textbf{in} $\tt color$) \\
\hspace*{2cm}         $\tt col[x,c] > 0 \Rightarrow 1 > 0$; \vspace*{2mm} \\
\hspace*{1cm}     \textbf{forall} ($\tt c$ \textbf{in} $\tt color$)  \\
\hspace*{2cm}        $\tt used\_color[c] > 0 \Leftrightarrow $ \textbf{sum}($\tt x$ \textbf{in} $\tt node$) $\tt col[x,c] > 0 ;$ 
\vspace*{2mm} \\
\hspace*{1cm}     \textbf{forall} ($\tt \<x,y\>$ \textbf{in} $\tt edge,\ c$ \textbf{in} $\tt color$)  \\
\hspace*{2cm}         $\tt col[x,c] * col[y,c] > 0 \Rightarrow \textbf{false}$; \\
\};
\hfill $\Box$
\end{quote}
\end{example}

\item \emph{Constraint optimization.} A very simple optimization
consists in deleting the OPL constraints whose head is always true
(e.g. the head consists of the constant $1$) as they are always
satisfied. For instance, in the above example the second OPL
constraint can be deleted as its head consists of the
constant~$1$.

An additional simple optimization can be performed by ``pushing
down" conditions defined inside the OPL constraints. For instance,
the following code:
\[
\begin{array}{l} {\small
\tt \textbf{Q}\ (X_1\ \textbf{in}\ D_1, \dots, X_k\ \textbf{in}\
D_k) \tt \ \ \ \<statement_1\>\ X_i \,\theta\,X_j \Rightarrow  \<statement_2\>
}\end{array}
\]
where $\textbf{Q}$ is either \textbf{forall} or \textbf{sum}, $\tt
X_i$ and $\tt X_j$ are either variables or constants or arithmetic expressions, $\theta$ is a
comparison operator,
can be rewritten as
\[
\begin{array}{l}{\small
\tt \textbf{Q}\ (X_1\ \textbf{in}\ D_1, \dots, X_k\ \textbf{in}\
D_k:\ X_i \,\theta\,X_j ) \tt \ \ \ \<statement_1\> \Rightarrow  \<statement_2\>
}\end{array}
\]

\item \emph{Arrays reduction.} A further optimization can be
performed by reducing the dimension of the arrays (of decision variables)
corresponding to some guess predicates. Specifically, given a
guess predicate $\tt s$ defined by generalized partition rules of
the form:
\[
\begin{array}{l}{\small
\tt \oplus_L s(X_1,\dots,X_k,L) \leftarrow
body(X_1,\dots,X_k,Y_1,\dots,Y_n), dom(L) }\end{array}
\]
instead of declaring a  $(k\+1)$-dimensional array of boolean decision variables, it
is possible to introduce a $k$-dimensional array $\tt s$ of integer decision variables
ranging in $\tt \{0, \dots, |dom|\}$ and map each value in $\tt dom$
to $\tt \{1, \dots, |dom|\}$ by means of a one-to-one function.
The meaning of $\tt s[X_1,\dots,X_k]=c$ is that if ${\tt c} \neq 0$ then the atom
$\tt s(X_1,\dots,X_k,c')$ is $true$, where $\tt c'$ is the value in $\tt dom$
corresponding to the integer $\tt c$; if $\tt c = 0$ then
the atom $\tt s(X_1,\dots,X_k,c')$ is $false$ for any value $\tt c'$ in $\tt dom$.
Clearly, to make consistent the OPL program, every instance of $\tt
s[X_1,\dots,X_k,C]$ must be substituted with $\tt
(s[X_1,\dots,X_k]==C)$ and in each \textbf{forall} or \textbf{sum}
statement containing variables ranging in $\tt dom$
the condition $\tt C \neq 0$ must be verified.

\begin{example}\label{Optimized-OPL-Min-COloring}
The application of the previous optimizations to the program of Example
\ref{Min-COloring-Opt1} gives the following OPL program:
\begin{quote}\small
\textbf{int} $\tt\ cardcolor = \textbf{card}(color);$ \\
\textbf{range} $\tt intcolor = 0\,\.\.\,cardcolor;$\\
\textbf{dvar\ int} $\tt\ col[node]\ \textbf{in}\ intcolor ;$ \\
\textbf{dvar\ boolean} $\tt\ used\_color[intcolor];$ \vspace*{2mm}\\
\textbf{minimize} \vspace*{1mm}\\
\hspace*{6mm}     \textbf{sum}($\tt c$ \textbf{in} $\tt\  intcolor: c \neq 0)\ used\_color[c];$ \vspace*{1mm}\\
\textbf{subject to} \{ \\
\hspace*{6mm}     \textbf{forall} ($\tt x$ \textbf{in} $\tt node$) \\
\hspace*{12mm}         $1 > 0$ \ $\Rightarrow \tt \textbf{sum}(c\ \textbf{in}\ intcolor: c \neq 0)\ (col[x]\!==c) > 0$; \vspace*{2mm} \\
\hspace*{6mm}     \textbf{forall} ($\tt c$ \textbf{in} $\tt\ intcolor : c \neq 0$)  \\
\hspace*{12mm}        $\tt used\_color[c] > 0 \Leftrightarrow $ \textbf{sum}($\tt x$ \textbf{in} $\tt node$) $\tt (col[x]\!==\!c)>0;$\vspace*{2mm}\\
\hspace*{6mm}     \textbf{forall} ($\tt \<x,y\>$ \textbf{in} $\tt edge,\ c$ \textbf{in} $\tt \ intcolor :c \neq 0$)  \\
\hspace*{12mm}         $\tt (col[x]\!==\!c) * (col[y]\!==\!c) >0\Rightarrow  \textbf{false}$; \\
\};\hfill $\Box$
\end{quote}
\end{example}

\emph{Variable deletion.} A further optimization regards the
deletion of unnecessary variables and the reduction of domains.
For instance, in the last constraint in the OPL program of the
previous example, the variable $\tt c$ can be deleted as it is
just used to define the matching between $\tt col[X]$ and $\tt
col[Y]$. Thus, this constraint can be rewritten as:
\begin{quote}\small
\hspace*{6mm}     \textbf{forall} ($\tt \<x,y\>$ \textbf{in} $\tt edge$)  \\
\hspace*{12mm}         $\tt (col[X]\!== col[Y]\!) >0\Rightarrow \textbf{false}$; \\
\end{quote}

\vspace*{-4mm}
Observe that, if the body of the partition rule only
contains  database domains, the integer decision variables
of the guess predicate  can range in the
set of integers $\tt \{1, \dots, |dom|\}$ as the head atom is true
for all possible values of its variables $\tt X_1, \dots, X_k$.
This means that under such circumstances, it is not
necessary to introduce the additional condition stating that the
value of the variable cannot be $0$.
Under this rewriting, the first constraint can be deleted as its head is
always satisfied.

\vspace{2mm}
The following example shows the final version of the min coloring
program, obtained by applying the optimizations above.

\begin{example}\label{Optimized-OPL-Min-COloring-v2}
Min Coloring \em
(optimized version).

\begin{quote}\small
\textbf{int} $\tt\ cardcolor = \textbf{card}(color);$ \\
\textbf{range} $\tt intcolor = 1\,\.\.\,cardcolor;$\\
\textbf{dvar\ int} $\tt\ col[node]\ \textbf{in}\ intcolor ;$ \\
\textbf{dvar\ boolean} $\tt\ used\_color[intcolor];$ \vspace*{2mm}\\
\textbf{minimize}  \vspace*{1mm}\\
\hspace*{6mm}     \textbf{sum}($\tt c$ \textbf{in} $\tt intcolor)\ used\_color[c];$ \vspace*{1mm}\\
\textbf{subject to} \texttt{ \{ } \\
\hspace*{6mm}     \textbf{forall} ($\tt c$ \textbf{in} $\tt intcolor$)  \\
\hspace*{12mm}        $\tt used\_color[c] > 0 \Leftrightarrow $ \textbf{sum}($\tt x$ \textbf{in} $\tt node$) $\tt (col[x]==c) > 0; $ \vspace*{1mm} \\
\hspace*{6mm}     \textbf{forall} ($\tt \<x,y\>$ \textbf{in} $\tt edge) $ \\
\hspace*{12mm}         $\tt (col[x]==col[y]) > 0 \Rightarrow \textbf{false}$; \\
\};
\end{quote}
\end{example}

The OPL program corresponding to the k-coloring problem consists
of only one constraint, namely the second OPL constraint in the
previous example.

\end{itemize}

\paragraph{Aggregates.}

The current version of the paper does not include aggregates,
although the language could be easily extended with stratified aggregates
which can be effortlessly translated into OPL programs.

Consider, for instance, a digraph stored by means of the two relations $\tt node$ and $\tt edge$ and the following logic rule with aggregates\footnote{
The syntax used refers to the proposal presented in \cite{Greco99}. }:
\[\tt
out(X,C) \leftarrow edge(X,Y), count((X),C)
\]
computing for each node $\tt X$ the number $\tt C$ of outgoing arcs.
Such a rule could be easily translated into the following OPL script code:
\[
\begin{array}{l}
\tt \{\textbf{string}\}\ node = \dots ;\\
\tt \textbf{tuple}\ edge \{\textbf{string}\ a;\ \textbf{string}\ b ;\} ;\\
\tt \{edge\}\ edges = \dots ; \\
{\tt \textbf{int}\ out[node];}\ \\ 
\tt \textbf{execute}\{\\
\tt \hspace*{0.7cm}\textbf{for}\ ( \textbf{var}\ e\ in\ edges) \\
\tt \hspace*{1.5cm}       out[e\.a] = out[e\.a] +  1;\\
\tt \};
\end{array}
\]

In this paper, we have not considered aggregates since we would like to
define more efficient translations which allow us to express and efficiently compute
greedy and dynamic programming algorithms.
In the literature, there have been several proposals to extend Datalog with aggregates.
For instance, the proposal of \cite{Greco99} allows us to write rules with
stratified aggregates and
evaluate programs so that the behavior of  dynamic programming  is
captured  (see also \cite{GreZan01} for greedy algorithms).

Consider the query $\tt \<SP,stc(X,Y,C)\>$ computing the shortest
paths of a weighted digraph, where $\tt SP$ consists of the
following rules:
\[
\begin{array}{l}
\tt stc(X,Y,C) \leftarrow tc(X,Y,C), min((X,Y),C). \\
\tt tc(X,Y,C) \leftarrow edge(X,Y,C). \\
\tt tc(X,Y,C) \leftarrow edge(X,Z,C1), tc(Z,Y,C2), C=C1+C2.
\end{array}
\]
and weights associated with arcs are positive integers. A standard
translation and execution has two main problems: i) the computation
is not efficient since for each pair of nodes all paths with
different weights are considered, and ii) if the graph is cyclic the
computation never terminates (or terminates with an error). Since
shortest paths can be obtained by considering other shortest paths,
an OPL Script computing them could be as follows:

\[
\small
\begin{array}{l}
\hspace*{0mm}\it //\ declarations \\
\tt \textbf{tuple}\ edge \{\textbf{string}\ a;\ \textbf{string}\ b ;\ \textbf{int}\ c; \} ;\\
\tt \{edge\}\ edges =   \dots ; \\
\tt \{\textbf{string}\}\ node = \dots ; \\

\hspace*{0mm}\tt \textbf{int}\ tc\,[x\ \textbf{in}\ node][y\ \textbf{in}\ node]=\textbf{maxint}; \\
\hspace*{0mm}\tt \textbf{int}\ stc\,[x\ \textbf{in}\ node][y\ \textbf{in}\ node]=\textbf{maxint}; \vspace*{2mm} \\
\hspace*{00mm}\tt    \textbf{execute}\ \{\\
\hspace*{05mm}\it    //\ exit\ rule \\
\hspace*{05mm}\tt    \textbf{for}\ (\textbf{var}\ e_1\ \textbf{in}\ edge)\ \{ \\
\hspace*{10mm}\tt       tc[e_1\.a][e_1\.b]=e_1\.c;\\
\hspace*{10mm}\tt       stc[e_1\.a][e_1\.b]=e_1\.c;\\
\hspace*{05mm}\tt    \} \vspace*{2mm} \\
\end{array}
\]
\vspace*{-5mm}
\[
\small
\begin{array}{l}
\hspace*{05mm}\it    //\ recursive\ rule \\
\hspace*{05mm}\tt    \textbf{var}\ modified = \textbf{true}; \\
\hspace*{05mm}\tt    \textbf{while}\ (modified)\ \{ \\
\hspace*{10mm}\tt       modified = \textbf{false}; \\
\hspace*{10mm}\tt       \textbf{for}\ (\textbf{var}\ e\ \textbf{in}\ edges)\  \\
\hspace*{15mm}\tt              \textbf{for}\ (\textbf{var}\ y\ \textbf{in}\ node)\ \{\\
\hspace*{20mm}\tt                    tc[e\.a][y]=e\.c + stc[e\.b][y];\\
\hspace*{20mm}\tt                   \textbf{if}(tc[e\.a][y]<stc[e\.a][y])\ \{ \\
\hspace*{25mm}\tt                       modified=\textbf{true}; \\
\hspace*{25mm}\tt                       stc[e\.a][y]=tc[e\.a][y]; \\
\hspace*{20mm}\tt                   \} \\
\hspace*{15mm}\tt                \} \\
\hspace*{05mm}\tt    \} \\
\hspace*{0mm}\tt    \} \\
\end{array}
\]


\subsection*{Implementation and experiments}

\begin{figure}[h]
    \centering
    \vspace*{-0,5cm}
    \includegraphics[width=10cm]{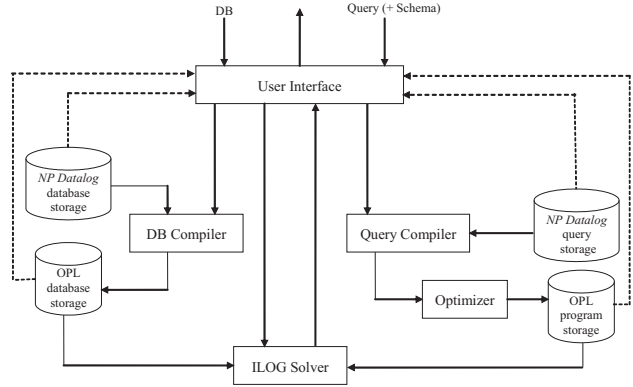}
    \vspace*{-1,5mm}
    \caption{System Architecture.}\label{fig-architecture}
\end{figure}

A system prototype translating \NPDA\ queries into OPL programs
and executing the target code using the ILOG OPL Development Studio
has been implemented.
The system architecture, depicted in Fig. \ref{fig-architecture},
consists of five main modules whose functionalities are next
briefly discussed.

\begin{itemize}
\item
\emph{User Interface} -- This module receives in input a
pair of strings identifying the file containing the source
database and the file containing the query.
If both the database and the query have already been translated, then the UI
asks the module \emph{ILOG Solver} to execute the query. If
the database (resp. query) has not been translated, then the UI sends
the name of the file containing the source database (resp. query) to the
module \emph{Database Compiler} (resp. \emph{Query Compiler}) to
be translated. Moreover, this module is in charge of visualizing
the answer to the input query.
\item
\emph{Database compiler} --
This module translates the source database into an OPL database.
\item
\emph{Query compiler} -- This module receives in
input an \NPDA\ query and gives in output
the corresponding OPL code. In order to check
the correctness of the query and generate the target code, the
module uses information on the schema of predicates.
\item
\emph{Optimizer} -- This module rewrites the OPL code
received from the module \emph{Query Compiler} and gives in output
the target (optimized) OPL code.
\item
\emph{Query executor} -- This module consists of the
ILOG OPL Development Studio which executes the query
stored by the module
\emph{Optimizer} into the \emph{OPL program storage}, over a
database stored into the \emph{OPL database storage}. The module
\emph{Query executor}  interacts with the module \emph{User
Interface} by providing it the obtained result.
\end{itemize}
Therefore, \NPDA\ can be also used to define a logic interface for
constraint programming solvers such as ILOG.
The experiments presented in this subsection show that the combination of the two
components is effective so that constraint solvers (as well as SAT solvers) can
be used as an efficient tool for computing logic queries whose semantics is based on
stable models.

In order to assess the efficiency of our approach,
we have performed several experiments comparing the performance obtained by implementing
\NPDA\ over the ILOG OPL Development Studio against
Answer Set Programming systems.
Specifically,\  \NPDA/OPL \ has been compared with
\emph{DLV}, \emph{Smodels}, \emph{ASSAT}, Clasp 
and \emph{XSB}.
The following version of the aforementioned
systems have been used:
\begin{itemize}
\item
ILOG OPL Development Studio 6.1
\cite{Ilog}
\item
DLV release 2007-10-11
\cite{dlvSys}
\item
Smodels 2.33  (and \emph{lparse} 1.1.1)
\cite{smodelsSys}
\item
ASSAT 2.02 (\emph{lparse} 1.1.1 and \emph{zChaff} 2007.3.12)
\cite{assatSys,zChaff}
\item
Clasp 1.2.1  (and \emph{lparse} 1.1.1)
\cite{claspSys}
\item XSB version 3.2 March 15, 2009
\cite{XSB}
\end{itemize}

The performances of the systems have been evaluated by measuring
the time necessary to find one solution of the following problems: 3-Coloring,
Hamiltonian Cycle, Transitive Closure, Min Coloring, N-Queens and Latin Squares.

For each system, we have used efficient encodings of the problems
which exploit efficient built-in constructs provided by the systems.
Every encoding and database used in the experiments can be downloaded
from the \NPDA\ web site
({\tt http://wwwinfo.deis.unical.it/npdatalog/}).

All the experiments were carried out on a PC with a processor Intel Core Duo 1.66 GHz and
1 GB of RAM under the Linux operating system .
%
In the sequel of this section the experimental results are presented.

\begin{figure}[h]
    \centering
    \begin{tabular}{ccc}
         \includegraphics[width=6cm, height=4cm]{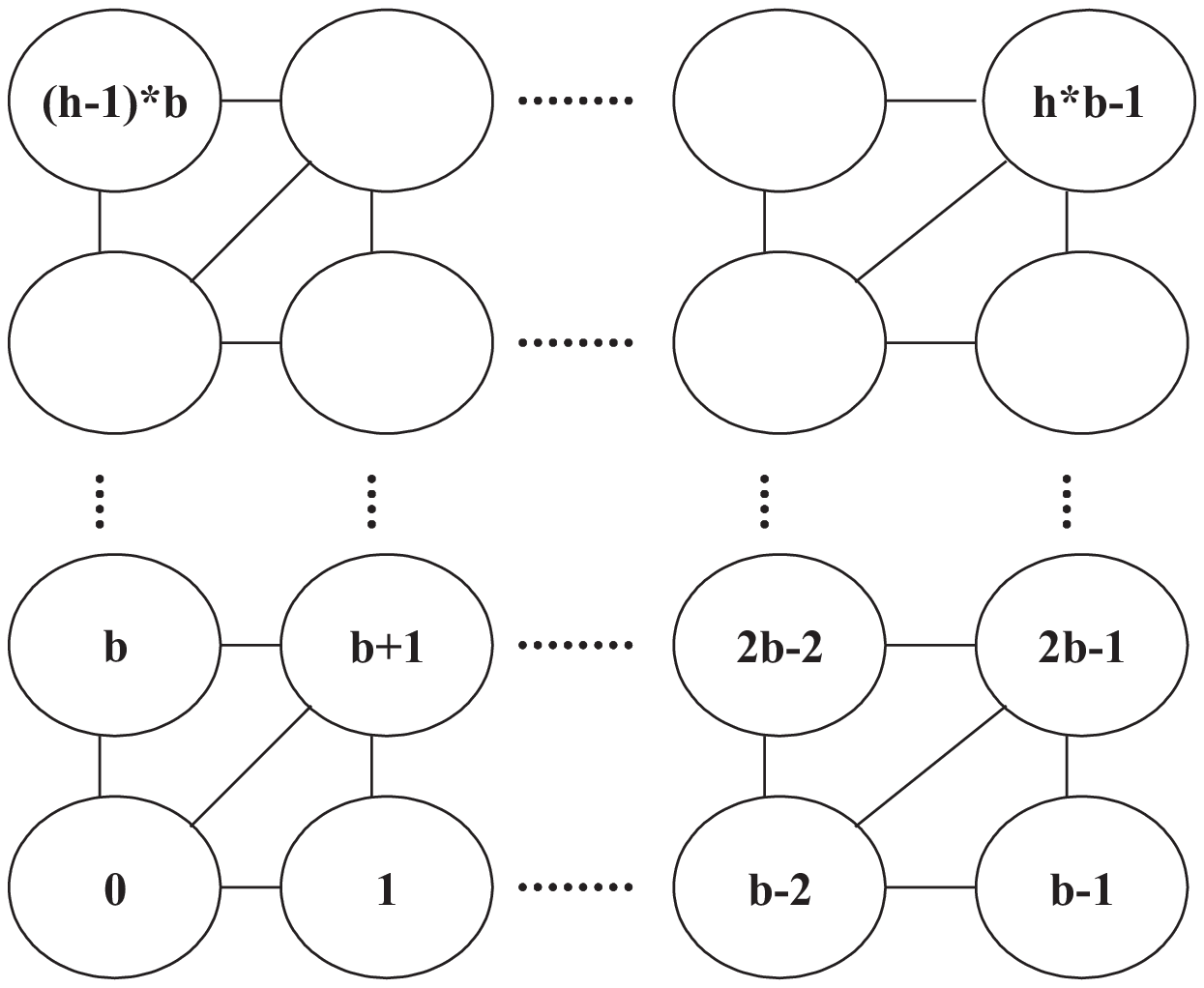} & &
         \includegraphics[width=6cm, height=4cm]{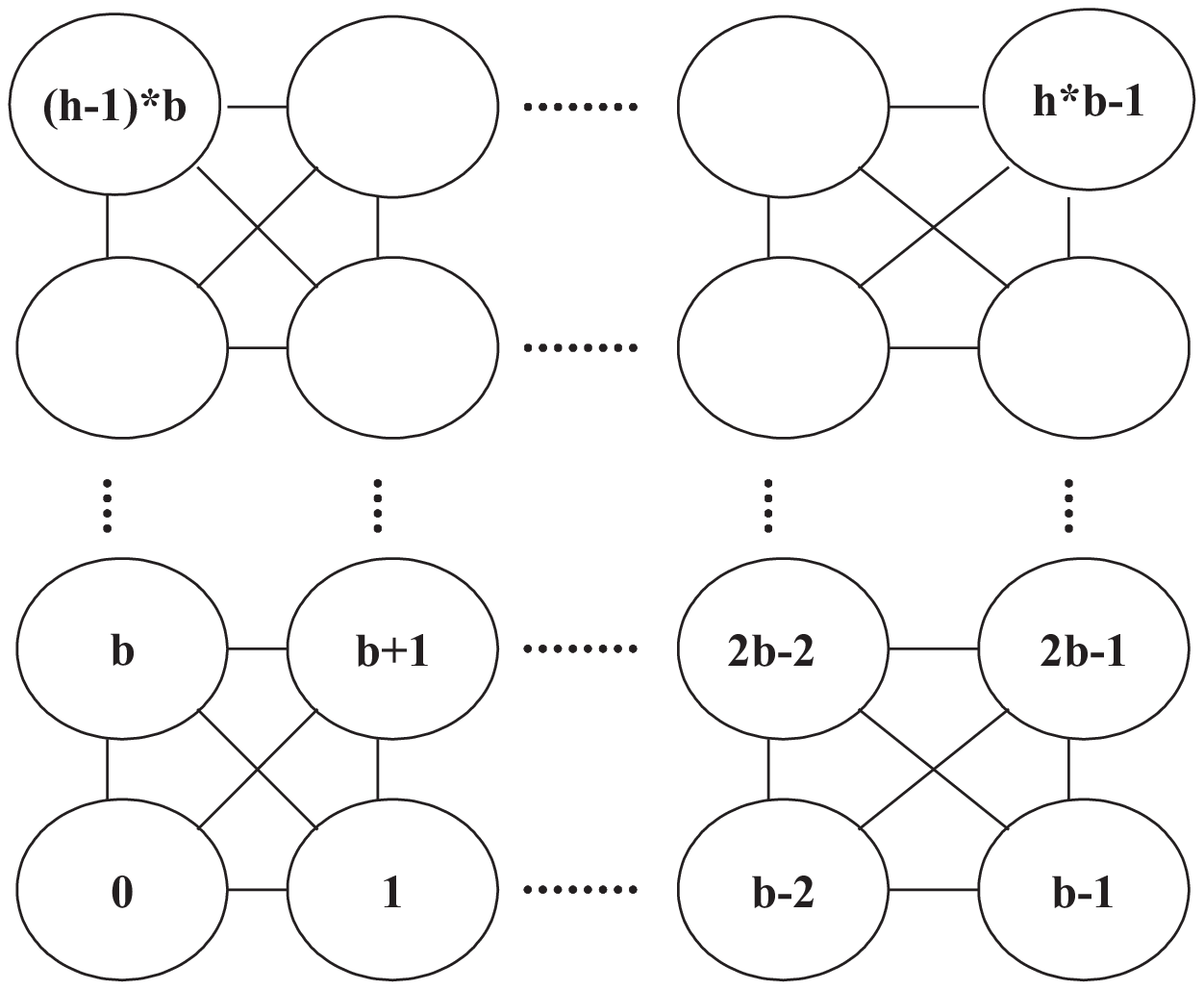}        \\
    (i) & & (ii)
    \end{tabular}
    \caption{Structured Graphs. \ \ \ \ \ \ \ \ \ }\label{Graph-Structure1}
\end{figure}

\begin{figure}[h]
 \hspace{-11.8mm}
 \begin{minipage}[c] {5cm}
    \centering
    \includegraphics[width=6.2cm,height=4.5cm]{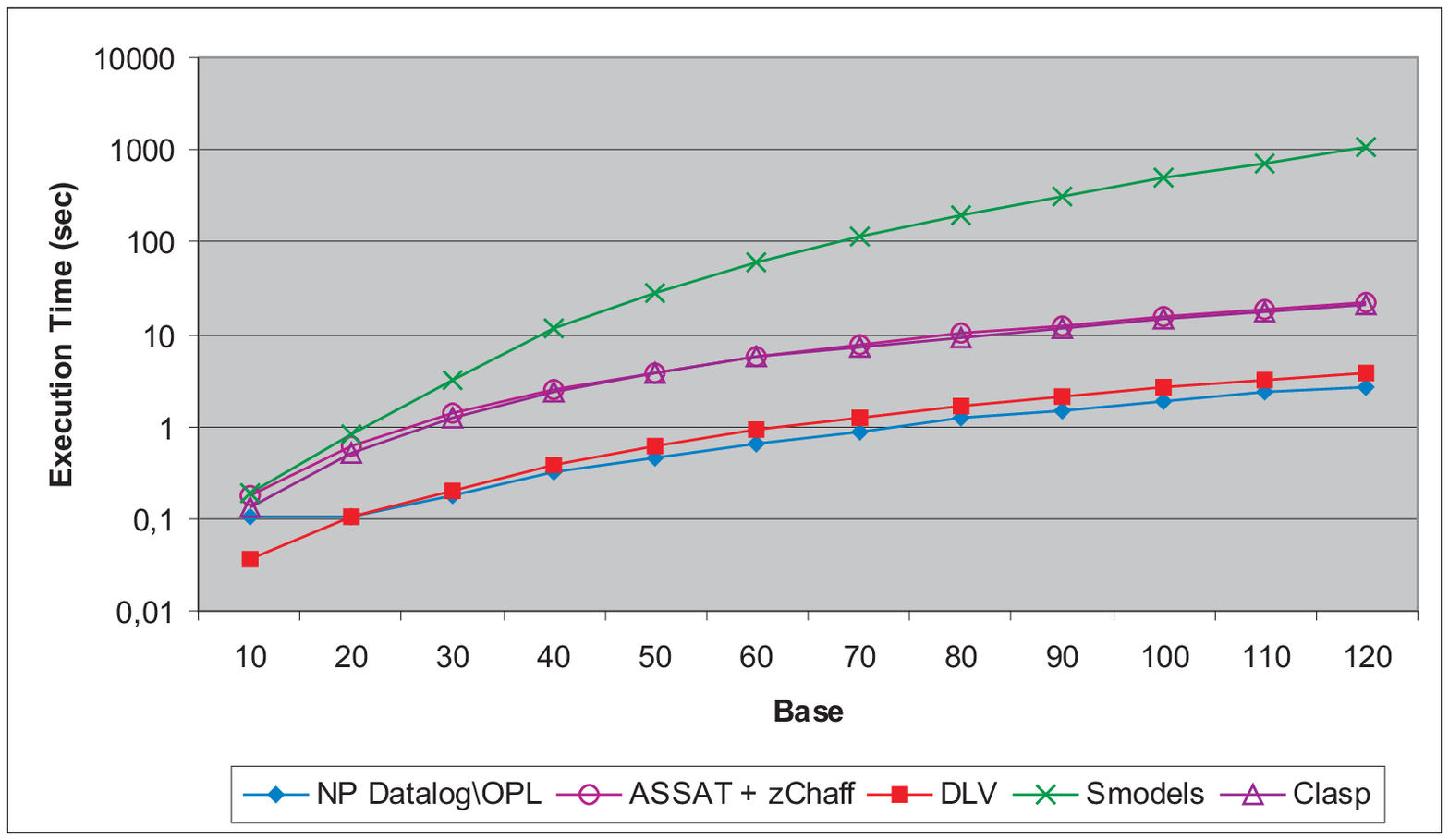}
    \caption{Execution time for the 3-coloring problem on structured graphs.}\label{fig:3ColResults-GR1}
 \end{minipage}
 \ \hspace{10mm}  \
 \begin{minipage}[c] {5cm}
    \centering
    \includegraphics[width=6.2cm,height=4.5cm]{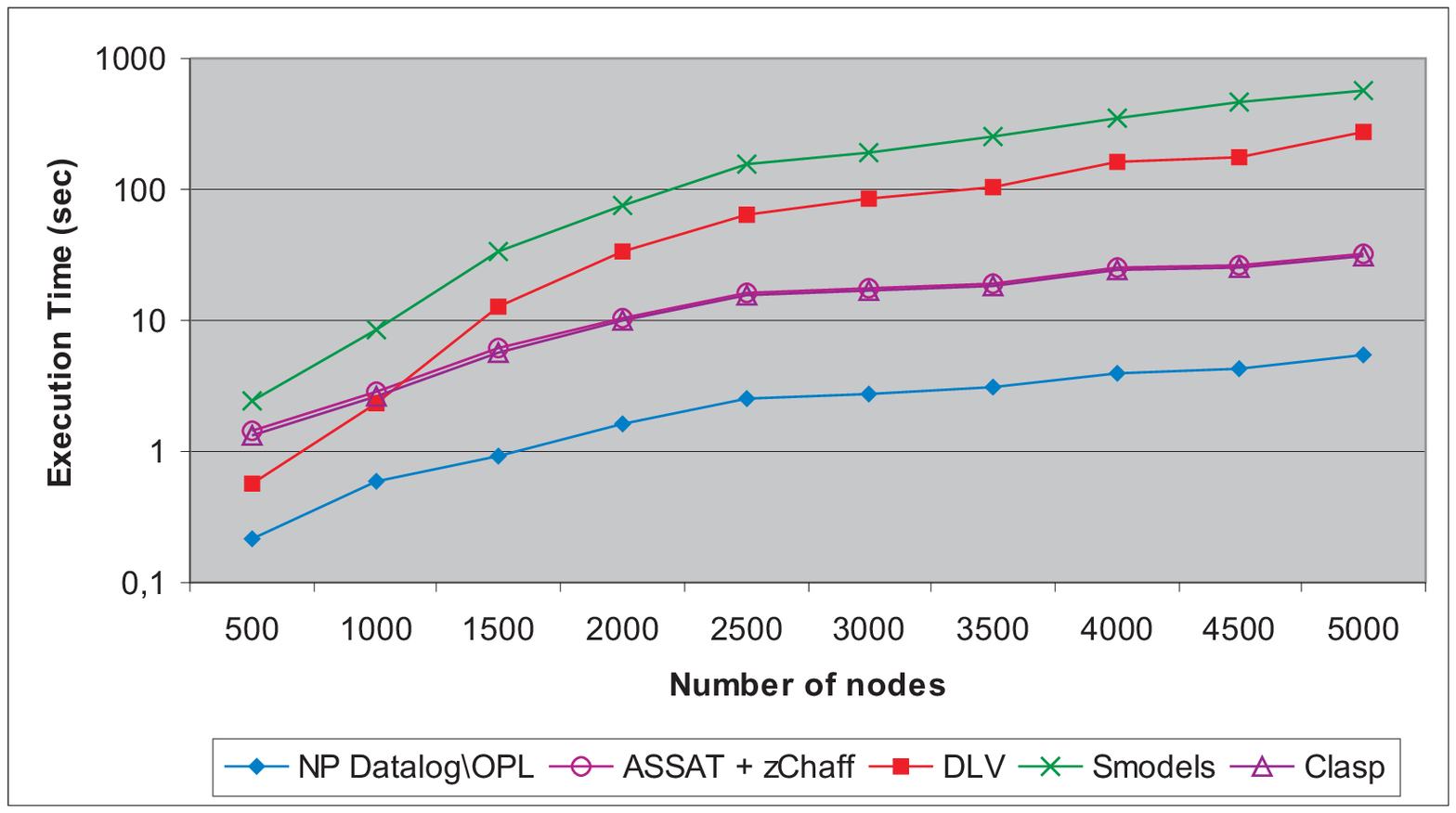}
    \caption{Execution time \  for \ the 3-coloring problem on random graphs.}\label{fig:3ColResults-GRandom}
 \end{minipage}
\end{figure}

\vspace*{-2mm}
\paragraph{3 Coloring.}
The 3-Coloring query has been evaluated on structured graphs of the form
reported in Fig.~\ref{Graph-Structure1}(i) and random graphs.
Specifically, structured graphs with $base = height$ have been used
(here $base$ denotes the number of nodes in the same row, $height$ the number of nodes in the same column;
the total number of nodes in the graph is $base * height$).
The random graphs have been generated  by means of  Culberson's graph generator~\cite{Culb1}.
Specifically, the following parameters have been used:
\emph{K-coloring scheme} equal to \emph{Equi-partitioned},
\emph{Partion number} equal to 3,
\emph{Graph type} is \emph{IID (independent random edge assignment)}.
Both structured and random graphs are all 3-colorable;
the results, showing the
execution times (in seconds) as the size of the graph increases, are reported in Fig.~\ref{fig:3ColResults-GR1}
and Fig.~\ref{fig:3ColResults-GRandom}, respectively.

As for structured graphs, the $x$-axis reports the number of
nodes in the same layer (i.e. the value of $base$).
\NPDA\ and DLV are faster than the other systems; ASSAT and Clasp have almost the same execution times
(observe that the scale of the $y$-axis is logarithmic).

Regarding random graphs, it is worth noting that we have considered, for
each number of nodes, five different graphs.
Thus, the execution times reported in Fig.~\ref{fig:3ColResults-GRandom} have been
obtained by evaluating the query five times
(over different graphs with the same number of nodes) and computing the mean value.
\NPDA/OPL is faster than the other systems; again, ASSAT and Clasp have almost the same execution times.

\paragraph{Hamiltonian Cycle.}

\begin{figure}[h]
 \hspace{-11.8mm}
 \begin{minipage}[c] {5cm}
    \centering
    \includegraphics[width=6.2cm,height=4.5cm]{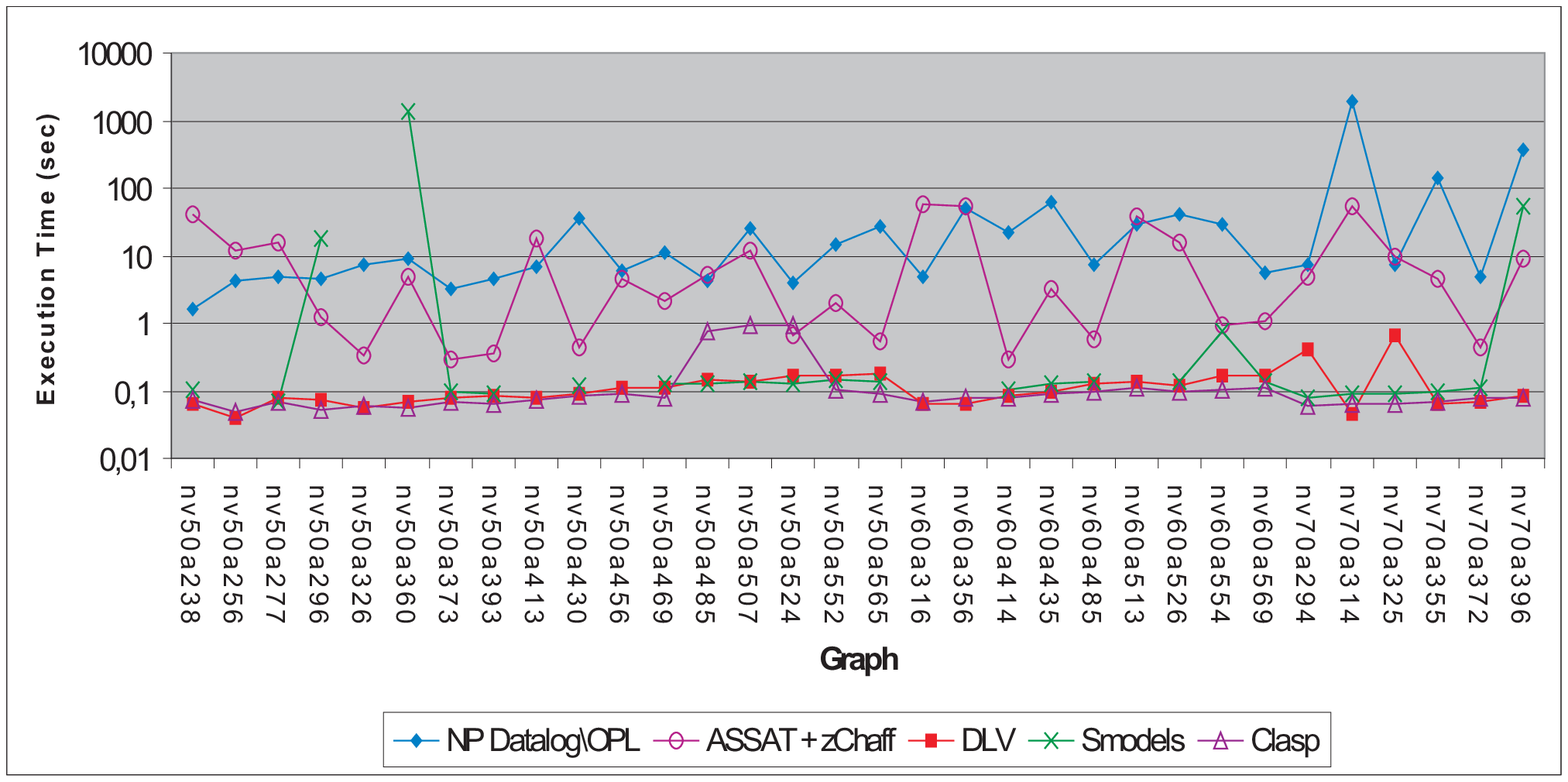}
    \caption{Execution time for the Hamiltonian Cycle problem on benchmark graphs.}\label{fig:HC-ASSAT}
 \end{minipage}
 \ \hspace{10mm}  \
 \begin{minipage}[c] {5cm}
    \centering
    \includegraphics[width=6.2cm,height=4.5cm]{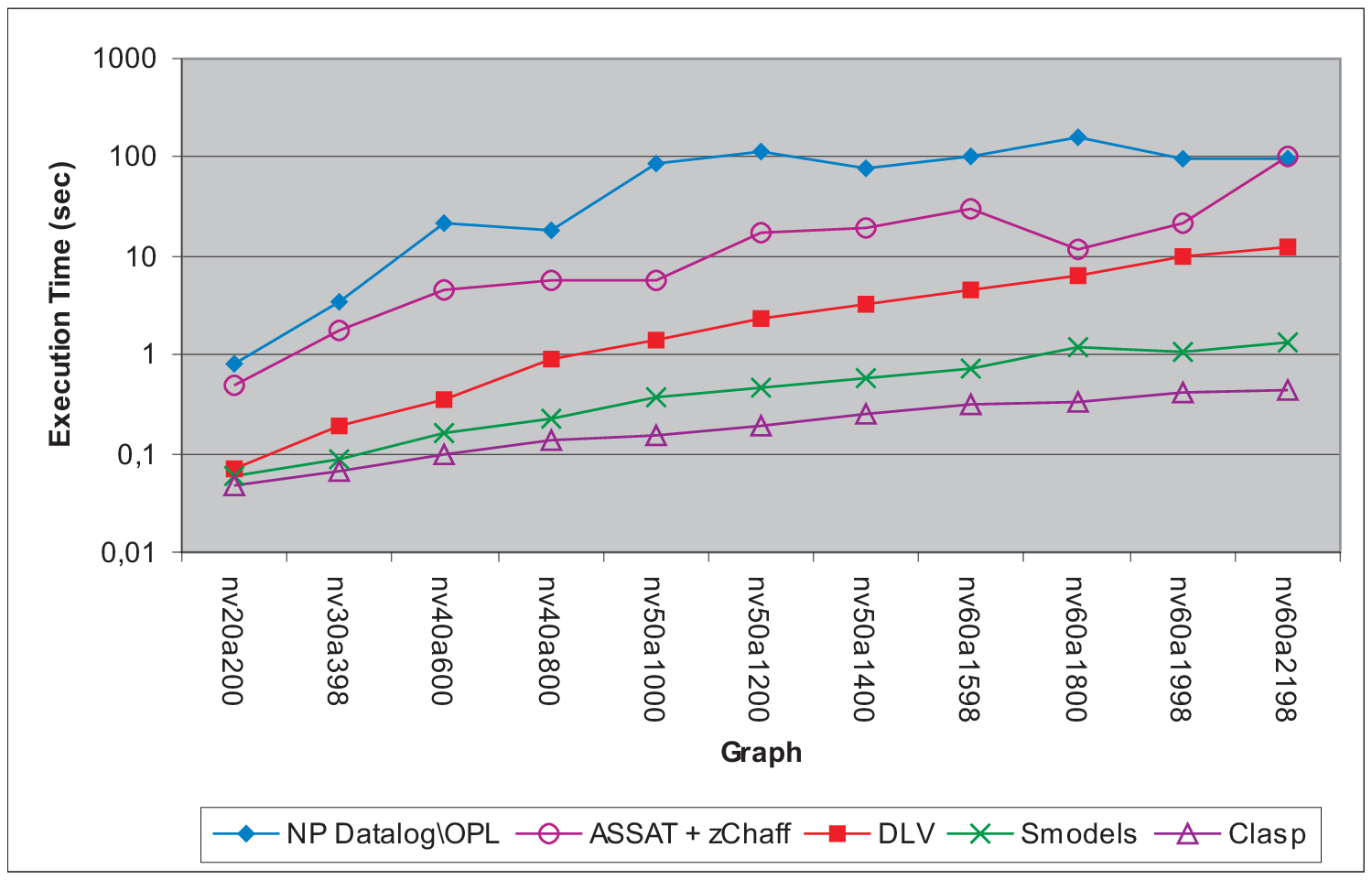}
    \caption{Execution time for the Hamiltonian Cycle problem on random graphs.}\label{fig:HC-random}
 \end{minipage}
\end{figure}

The Hamiltonian Cycle problem has been evaluated over benchmark graphs used to test other systems~\cite{HC-ASSAT-instances}
and random graphs generated by means of Culberson's graph generator~\cite{Culb2}.
All the graphs have a Hamiltonian cycle.
The \NPDA\ encoding (as well as the encodings for the other systems) can be found on~\cite{NPDatalogweb}.
The results are reported in Fig.~\ref{fig:HC-ASSAT} and Fig.~\ref{fig:HC-random}.
The $x$-axis reports the used graphs: a label $nvXaY$ refers to a graph with $X$
nodes and $Y$ arcs.
Observe that, in Fig.~\ref{fig:HC-ASSAT}, a missing value means that the system
has not answered in 30 minutes.
Clasp is the fastest system for both types of graphs.
DLV and Smodels are on average faster than the remaining systems.
For large ``dense'' graphs Smodels outperforms DLV, but
on some benchmark instances it runs out of time.

\paragraph{Transitive Closure.}

\begin{figure}[h]
    \centering
    \includegraphics[width=6.2cm,height=4.5cm]{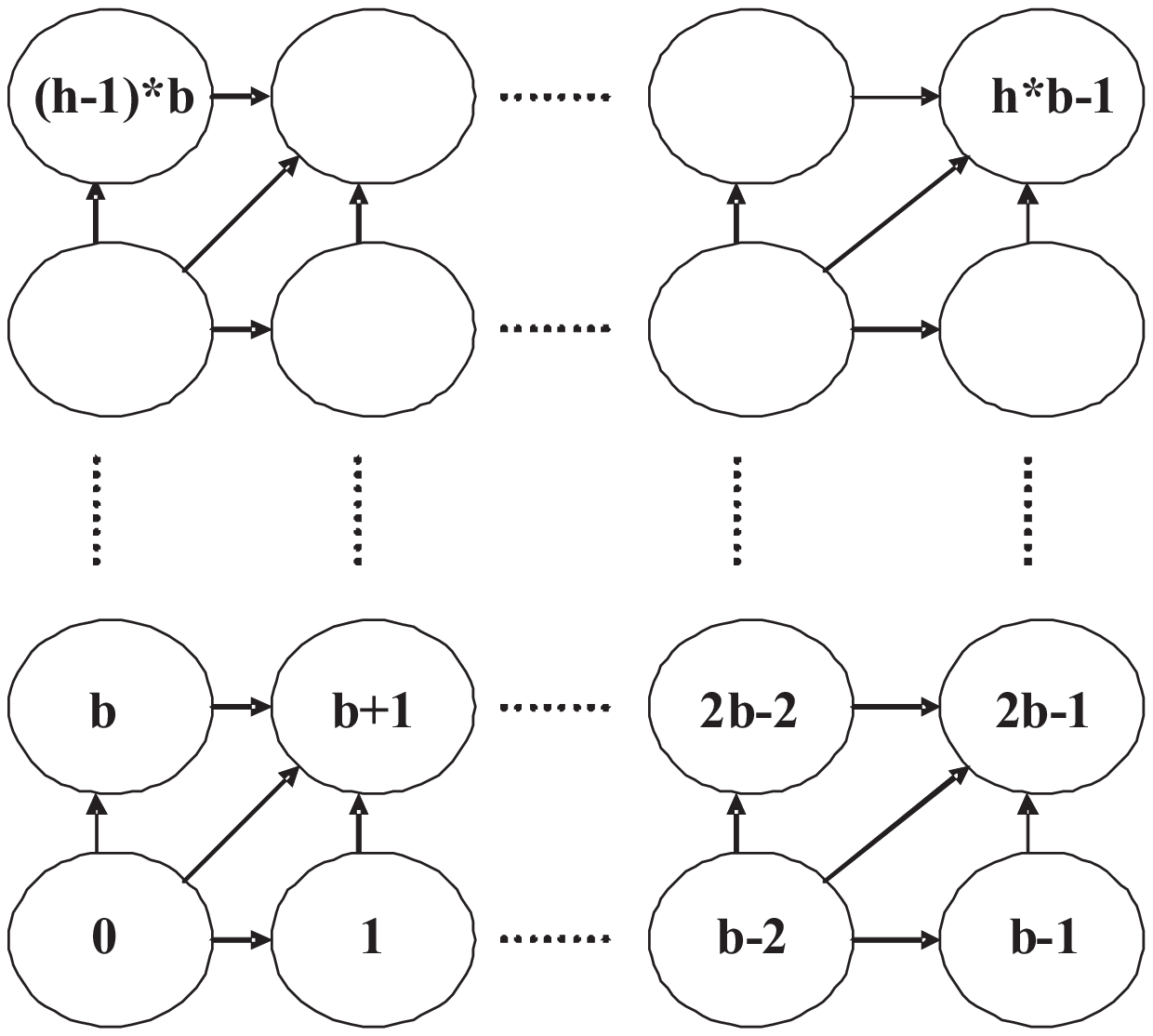}
    \caption{Directed structured graphs.}\label{fig:directed-structured}
\end{figure}

The Transitive Closure problem has been evaluated over directed structured graphs such as those reported in
Fig.~\ref{fig:directed-structured}.
Specifically, instances with $base = height$ have been used
($base$ denotes the number of nodes in the same row, $height$ the number of nodes in the same column).

\begin{figure}[h]
    \centering
    \includegraphics[width=6.2cm,height=4.5cm]{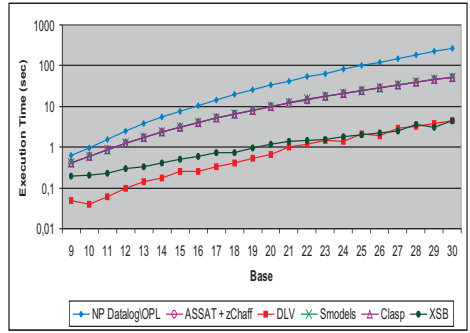}
    \caption{Execution time for the Transitive Closure problem on directed structured graphs.}\label{fig:TC-ASSAT}
\end{figure}

The results, which are reported in Fig.~\ref{fig:TC-ASSAT},
show that DLV and XSB are faster than the other systems; ASSAT, Clasp and Smodels
almost have the same execution times.

\paragraph{Min Coloring.}

\begin{figure}[h]
    \centering
    \begin{tabular}{ccc}
        \hspace{-0.1cm}\includegraphics[width=6.2cm, height=4.5cm]{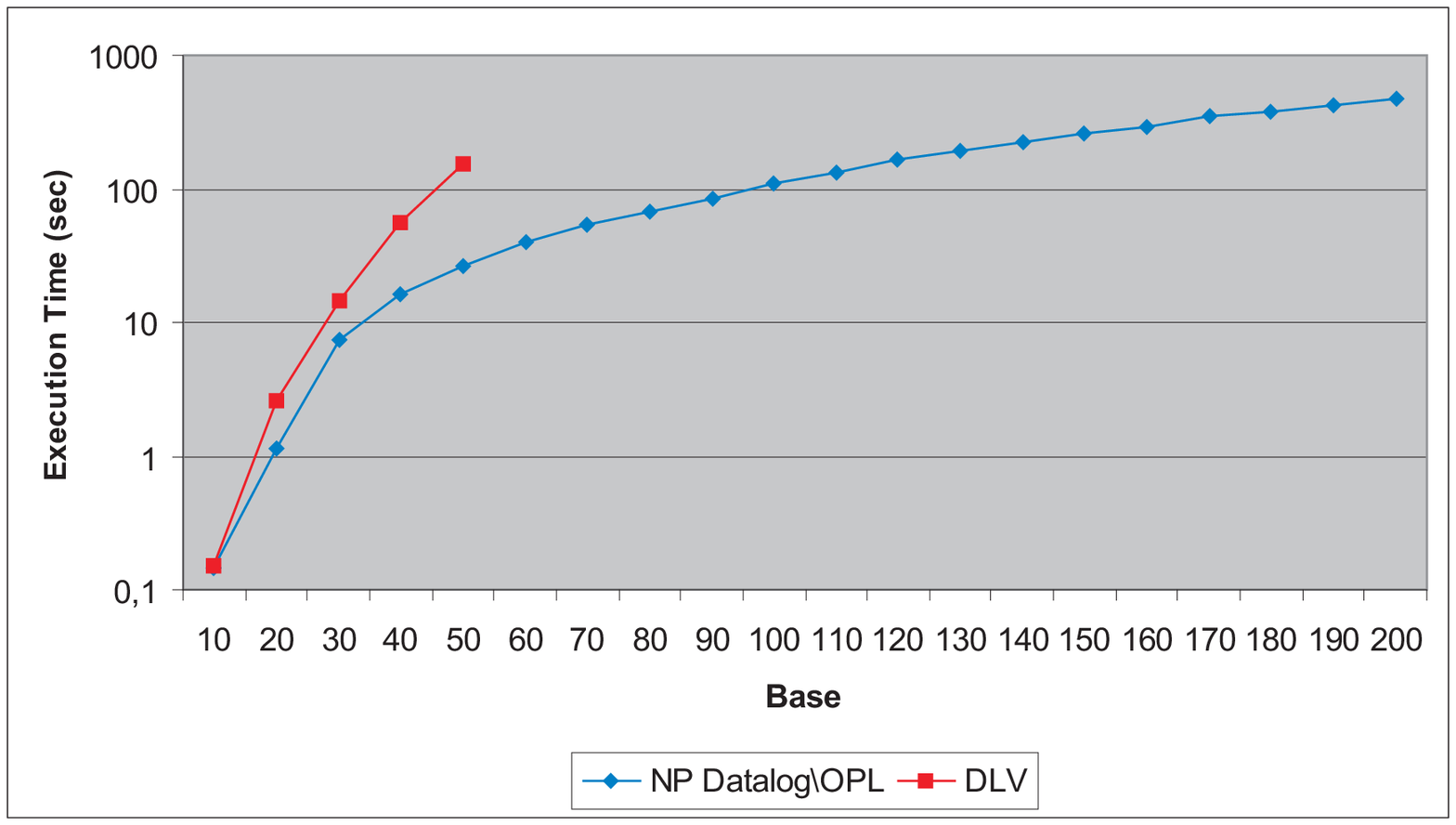} & &
        \hspace{-0.6cm}\includegraphics[width=6.2cm, height=4.5cm]{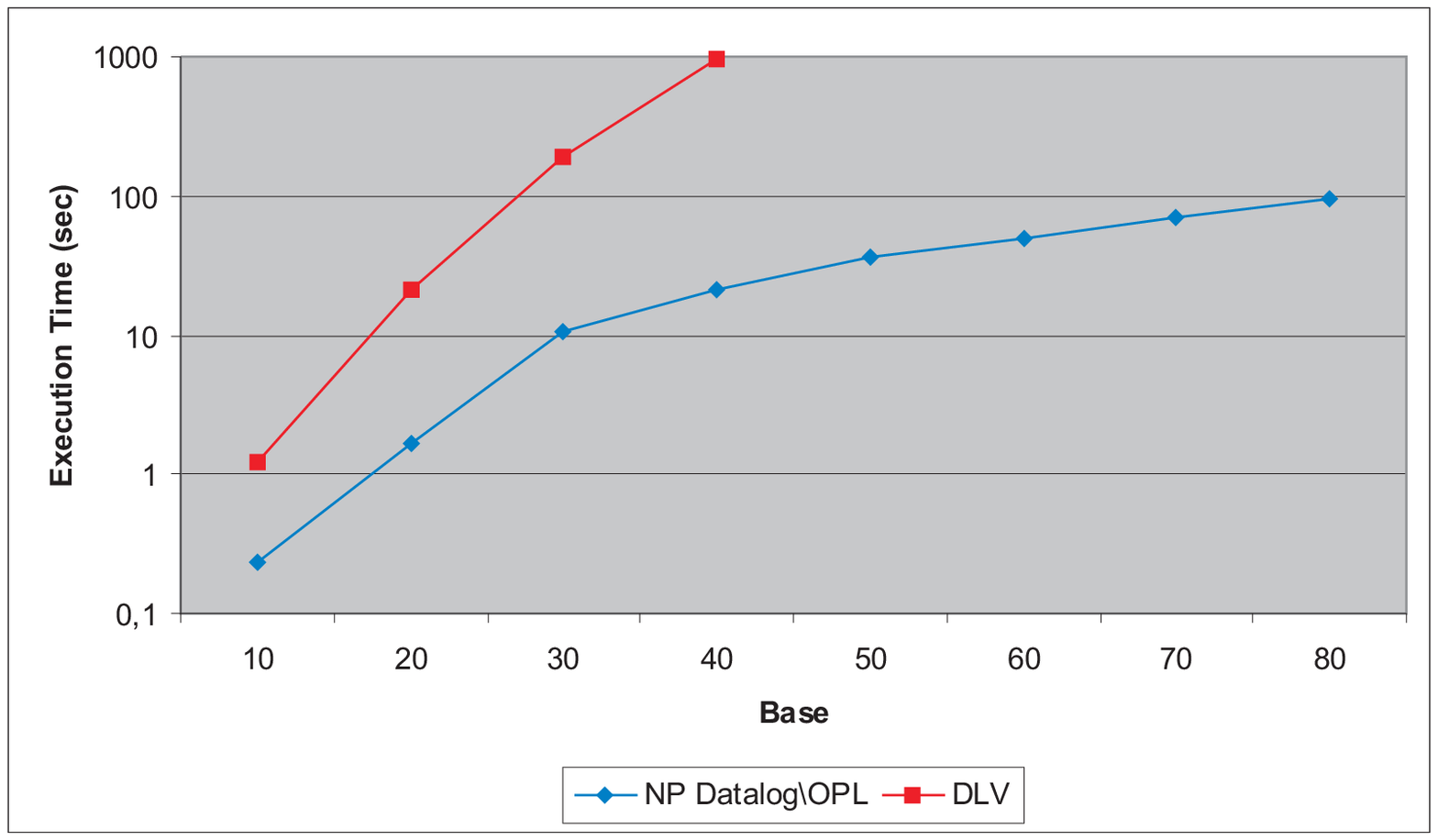} \\
    (i) & & (ii)
    \end{tabular}
    \caption{Execution time for the Min Coloring problem on structured graphs.}\label{fig:MinColResults}
\end{figure}

As for the Min Coloring optimization problem, we have used structured graphs such as those of Fig.~\ref{Graph-Structure1}.
Instances having the structure reported in Fig.~\ref{Graph-Structure1}(i)
need at least three colors to be colored, whereas instances having
the structure reported in Fig.~\ref{Graph-Structure1}(ii) need at least four
colors to be colored.
The number of colors available in the
database has been fixed for the two structures, respectively, to
four and five (one more than the number of colors necessary to color
the graph).
The results are reported in Fig.~\ref{fig:MinColResults} and show that
\NPDA\ outperforms DLV.
A missing time means that DLV runs out of time
(also in this case we had a 30 minute time-limit).

\paragraph{N-Queens.}

\begin{figure}[h]
    \centering
    \includegraphics[width=6.2cm,height=4.5cm]{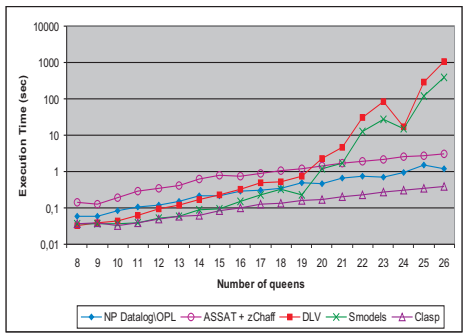}
    \caption{Execution time for the N-Queens problem.}\label{fig:Queens}
\end{figure}

We have considered empty chessboards to be filled with $N$ Queens for increasing values of $N$.
The results are reported in Fig.~\ref{fig:Queens}.
Clasp is faster than \NPDA\ which is in turn faster than ASSAT;
for a high number of queens, DLV and Smodels become slower than the other systems.

\paragraph{Latin Squares.}

\begin{figure}[h]
    \centering
    \includegraphics[width=6.2cm,height=4.5cm]{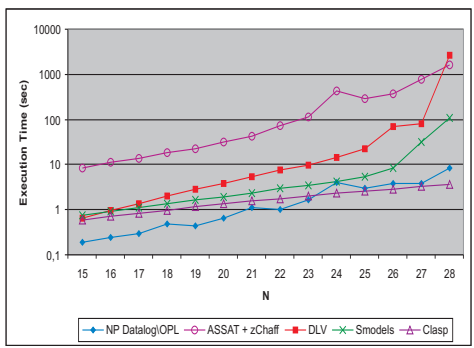}
    \caption{Execution time for the Latin Squares problem on random squares.}\label{fig:LS}
\end{figure}

We have considered partially filled tables which have been generated randomly.
In every table, $60\%$ of the squares are empty.
We have considered, for each table size, five different instances.
Thus, the execution times reported in Fig.~\ref{fig:LS} have been obtained by evaluating the query five times
(over different tables of the same size) and computing the mean value.
The results show that \NPDA\ and Clasp are faster than the other systems.\\

The experimental results reported above show that our system only seems to suffer
with programs where the evaluation of the deterministic components is predominant.
The reason is that deterministic components (often consisting of recursive rules)
are translated into OPL scripts, which correspond to the evaluation of
such components by means of the naive fixpoint algorithm,
whereas problems which can be expressed without recursion (or in which the non-deterministic
components are predominant) are executed efficiently.
The implementation of our system prototype could be enhanced by making
more efficient the translation of stratified (sub)programs or by using a different
evaluator for these components.
For instance, they could be evaluated by means of ASP systems, thus combining
their efficiency in the computation of deterministic components with the efficiency
of OPL in the computation of non-deterministic components.

\section{Related Languages and Systems}

Several languages have been proposed for solving \NP\ problems.
Here we have analyzed three different classes of
languages: specification languages, constraint and logic programming
languages, and answer set logic languages.

\vspace{-1mm}
\subsection*{Specification Languages}

Specification languages  are highly declarative and allow the user
to specify problems in terms of guess and check techniques.

\begin{description}
\item
\texttt{NP-SPEC }\cite{CadIan01,CadIan05} is a logic-based
specification language allowing the built-in second-order
predicates {\tt Subset, Partition, Permutation} and {\tt
IntFunc}. The semantics of an \texttt{NP-SPEC} program is based on
the notion of model minimality and the language upon which this
semantics relies on is $DATALOG^{CIRC}$, i.e. an extension of \DA\
in which only some predicates  are minimized and the
interpretation of the other is left open. An
\texttt{NP-SPEC} program consists of two sections: the DATABASE
section, specifying the instance and  the SPECIFICATION section
specifying the question.
To make NP-SPEC  executable, specifications are translated into
SAT instances and then executed using a SAT solver.

\item
\emph{KIDS} (Kestrel Interactive Development System) \cite{kids}
is a semi-automatic program development system  that, starting
from an initial specification of the problem, produces an
executable code through a set of consistency-preserving
transformations. The problem is written in a logic based language
augmented with set-theoretic data types and functional constraints
on the input/output behavior. To make the language executable,
specifications are firstly translated into $Common Lisp$ and then
into machine code. Before the compilation task, the user may
select an optimization technique, such as simplification or
partial evaluation, to obtain a more efficient target code.

\item \emph{SPILL-2}
(SPecifications In a Logic Language) is the second version of an
executable typed logic language that is an extension of the
Prolog-like language Goedel \cite{Spill97}. A specification in
\texttt{SPILL-2} consists of a set of type declarations, a set of
function declarations,  a set of predicate declarations and a
number of logical expressions (queries) that are used to test the
specification. A specification in \texttt{SPILL}  is required to
be ``executable" in the sense that it is possible to ``test"
whether a  provided solution is feasible w.r.t. a given
specification. The execution of a program consists in evaluating
each query in the context of the specification and reporting the
result (\emph{true} if the query succeeds and \emph{false}
otherwise).

\end{description}

\subsection*{Constraint and Logic Programming Languages}\label{Related-Work}

The basic idea of constraint programming (CP) is to model and
solve a problem by exploring the set of constraints that fully
characterize the problem. Almost all computationally hard
problems, such as planning, scheduling and graph theoretic
problems, fall into this category. A large number of systems (more
than 40) for solving CP problems have been developed in computer
science and artificial intelligence:

\begin{itemize}

\item
\emph{Constraint Logic Programming}, an extension of logic
programming able to manage constraints, started about 20 years ago
by Jaffar et~al. \cite{CLP92}. Several constraint logic languages
allowing the formulation of constraints over different domains
exist. Basically, all these languages embed efficient constraint
solvers in logic based programming languages, such as
Prolog. Here we cite, among the others, CLP \cite{MarStu99},
SICStus Prolog \cite{sicstusSys}, BProlog \cite{bprolog}, ECLiPSe
\cite{WalSch*99} and Mozart \cite{VanRoy99}.
%
%

\item
\emph{ILOG OPL Development Studio} \cite{Ilog}, an integrated development
environment for mathematical programming and combinatorial
optimization applications. The syntax of OPL is well-suited to
express optimization problems defined in the mathematical
programming style \cite{VAN99,VAN99b}.

\item
\emph{Constraint LINGO} \cite{Finkel*04}, a high-level
logic-programming language for expressing tabular
constraint-satisfaction problems such as those found in logic
puzzles and combinatorial problems such as graph coloring.

\end{itemize}

Several languages extending Prolog have been proposed as well.
Most of these languages have been designed to provide powerful capabilities
to represent and solve general problems and not to solve \NP\ problems.
Here we mention:
\begin{description}
\item
\emph{BinProlog} \cite{BinProlog}, a fast and compact Prolog compiler, based on the transformation
of Prolog to binary clauses.
BinProlog is based on the BinWAM abstract machine, a specialization of the
WAM for the efficient execution of binary logic programs.
\item
XSB \cite{Warren-XSB}, an extension of Prolog supporting the well-founded
semantics \cite{VanGelder} and including implementations
of OLDT (tabling) and HiLog terms. OLDT resolution is extremely useful
for recursive query computation, allowing programs to terminate correctly
in many cases where Prolog does not. HiLog supports a type of higher-order
programming in which predicate symbols can be variable or structured.
\end{description}

An extension of classical first order logic, called \emph{ID-Logic}, has been
proposed in \cite{Den00}.
Basically, in an ID-Logic theory, we can distinguish four different components describing
i) data, ii) open predicates, iii) definitions and iv) assertions (or constraints).
The relationships between ID-Logic and ASP has been studied in \cite{MarGil*04},
where it has been also presented how ID-Logic theories can be translated into
\DAm\ programs under ASP semantics.

\subsection*{Answer Set Programming Languages and Systems}

Several deductive systems based on stable model semantics have
been developed too. Here we discuss some of the more interesting answer-set based systems
and languages:

\begin{description}
\item
DLV (Vienna Univ. of Technology and University of Calabria)  \cite{dlv,leone-sys}  is
a deductive database system, based on disjunctive logic
programming. DLV extends Datalog with general
negation, inclusive head disjunction and two
different forms of constraints: strong constraints, which must be
satisfied, and weak constraints, which are satisfied if possible
(preferred models are those which minimize the number of ground
weak constraints which are not satisfied).
For instance, the program of Example 1 is a DLV program, whereas
by replacing exclusive disjunction with inclusive disjunction in
the program of Example 2 we get a DLV program (the minimality of the
models guarantees that every node cannot belong to both relations
$\tt v$ and $\tt nv$).
The optimization query of examples 1 and 2 can be defined by
adding the weak constraint $\tt :\sim v(X)$ which minimizes the
number of ground false weak constraints (i.e. $\tt v$-tuples).

\item
Smodels (Helsinki Univ. of Technology) \cite{smodels} is a
system for answer set programming consisting of Smodels,
an efficient implementation of  the stable model semantics for
normal logic programs and {\em lparse}, a front-end that
transforms user programs so that they can be understood by
Smodels.

Besides standard rules \emph{lparse} also supports a number of
extended rules: \emph{choice, constraint and weight rules}. The
formal semantics of all three types of rules can be defined through the use  of
\emph{weight constraints} and \emph{weight constraint rules}. In
\emph{lparse} the weight constraints are implemented as special
literal types. Basically, a weight constraint is of the form: {
\tt $L \leq l_1 =w_1,\dots, l_n= w_n \leq U$ } where
$l_1,\dots,l_n,$ are literals, $L$ and $U$ are the integral lower
and upper bounds, and $w_1,\dots, w_{n}$ are weights of the literals.
The intuitive semantics of a weight constraint is that it is
satisfied exactly when the sum of weights of satisfied literals
$l_1,\dots,l_n$  is between $L$ and $U$, inclusive. A weight
constraint rule is of the form $C_0 \leftarrow C_1,\dots,C_n$
where $C_0,\ldots,C_n$ are weight constraints. Besides the use of
literals, \emph{lparse} also enhances the use of \emph{conditional
literals} having the form: {\tt $ p(X) : q(X)$} where $p(X)$ is
any basic literal and $q(X)$ is a domain predicate.

\item
\emph{Datalog Constraint.}  \cite{EastTru00} proposed a
new nonmonotonic logic, called {\emph Datalog with constraints} or
\emph{DC}. A \emph{DC} theory consists of constraints and Horn
rules (Datalog program).  The language is determined by a set of
atoms $At = At_C \cup At_H$ where $At_C$ and $At_H$ are disjoint.
Formally, a \emph{DC} theory is a triple $T=(T_C, T_H, T_{PC})$
where $T_C$ is a set of constraints over $At_C$, $T_H$ is a set of
Horn rules whose head atoms belong to $At_H$ and $T_{PC}$ is a set
of constraints over $At$ (\emph{post constraints}).
The problem of the existence of
an answer set, for a finite propositional $DC$ theory $T$, is
$\NP$-complete \cite{EastTru00}.

\item
ASSAT. \cite{ASSAT} proposed a translation
from normal logic programs with constraints under the answer set
semantics to propositional logic. The peculiarity of this
technique consists in the fact that for each loop in the program, a
corresponding loop formula to the program's completion is added. The
result is a one-to-one correspondence between the answer sets of
the program and the models of the resulting propositional theory.
As in the worst case the number of loops in a logic program can
be exponential, the technique proposes to add a few loop formulas
at a time, selectively. Based on these results,
a system called ASSAT(X), depending on the SAT solver
X used, has been implemented for computing answer sets of a normal logic program
with constraints.

\item
Cmodels \cite{cmodels1,cmodels2} is an answer set programming
system that uses the frontend lparse and whose main computational
characteristic is that it computes answer sets using a
SAT solver for search.
Cmodels deals with programs that may contain disjunctive,
choice, cardinality and weight constraint rules.
The basic execution steps of the system can be outlined as follows:
(1) the program's completion is produced;
(2) a model of the completion is computed using a SAT solver;
(3) if the model is indeed an answer set, then the model is returned,
otherwise the system goes back to Step~2.
The idea is thus to use a SAT solver for generating
model candidates and then check if they are indeed the answer sets of a program.
The way Step 3 is implemented depends on the class of a logic program.

\item
Clasp \cite{clasp} is an answer set solver for (extended) normal logic programs.
It combines the high-level modeling capacities of answer set programming (ASP)
with state-of-the-art techniques from the area of Boolean constraint solving.
In fact, the primary Clasp algorithm relies on conflict-driven learning,
a technique that proved successful for satisfiability checking (SAT).
Unlike other ASP solvers that use conflict-driven learning,
Clasp does not rely on legacy software, such as a SAT solver or any other existing ASP solver.

\item
A-Prolog \cite{Gel01} is a logic language whose semantics is based on stable models,
designed to represent defaults (i.e. statements of the form ``Elements of a class $C$ normally
satisfy property $P$"), exceptions and causal effects of actions (``statement $F$ becomes
true as a result of performing an action $A$").
In the same work, an extension of the language, called \emph{ASET-Prolog}, is presented.
Such an extension enriches the language with two new types of atoms:
\emph{s-atoms}, which allows us to define subsets of relations,
and \emph{f-atoms}, which allows us to express constraints on the cardinality of sets.
An interesting application showing how
declarative programming in A-Prolog can be used to describe the dynamic behavior of
digital circuits is presented in \cite{Balduccini}.

\end{description}

\subsection*{Comparison with the other approaches proposed in the literature} \label{Comparison}

\NPDA\ is related i) to specification languages, for the style of
defining problems, ii) to answer set languages, for the syntax and
declarative semantics, and iii) to constraint programming.

\vspace*{-2mm}
\subsubsection*{Specification Languages}
The problem with specification languages is the tradeoff between
the expressiveness of the formal notation and its execution.
In general, specifications can be executed only by blind search
through the space of all proofs. A possible solution consists in
adding (to specifications) refinements which improve the execution, but the
result could be a longer specification, containing details and,
consequently, hard to understand.

\begin{description}
\item
\texttt{NP-SPEC} programs have a structure similar to the one of
\NPDA\ programs, although from the syntax point of view, the use of
meta-predicates, in some cases, does not make programs shorter and
more intuitive (see, for instance, the N-Queen problem reported in
\cite{CadIan05}). {\tt NP-SPEC} uses in addition to standard
Datalog rules, also meta-predicates and set operators, whereas
\NPDA\ uses only standard Datalog rules with shortcuts for limited
forms of (unstratified) negation. Moreover, although there is no
difference in expressivity, guesses in \texttt{NP-SPEC} are
defined over base relations, whereas in \NPDA\ they are defined over
general `deterministic' relations defined by stratified Datalog
programs. As a further difference, the partition mechanism is more
general and flexible in \NPDA\ w.r.t. {\tt NP-SPEC} as in the latter
the number of partitions is fixed. Concerning the semantics
aspects, the declarative semantics of \texttt{NP-SPEC} programs is
based on the notion of model minimality, whereas those of \NPDA\
is based on stable models.



\item
\emph{KIDS} results are ``sensitive" to the implementation issue.
Indeed, the \texttt{KIDS} system is semiautomatic: the user is
asked to interact with the system in order to transform high level
declarative specification into an efficient, correct and
executable program. Moreover, the complexity of the final
implementation in \texttt{KIDS} can result in dramatic
improvements if specialized techniques are used. On the other
hand, \NPDA\ is a fully declarative language whose execution
process is automatically optimized by the ILOG OPL Development Studio.

\item
\emph{SPILL-2} is not meant to use the specification of a problem
in order to compute a solution, but to test the specification
against some specific case, i.e. to verify whether a given
specification implies certain intended properties or, in other
words, if a specified property is consistent with the
specification. As for differences, specification and queries in
\texttt{SPILL} are compiled to Prolog, whereas our approach
introduces specifications using \NPDA\ and then performs the
translation of queries into OPL programs. Moreover, \NPDA\ is
based on stable model semantics, whereas \texttt{SPILL} uses  a
pure first order semantics, i.e. it does not include any form of
model minimization operations. As for a further difference, it is worth noting that
 \texttt{SPILL} does not provide a
characterization of its expressive power and its complexity.

\end{description}

\subsubsection*{Constraint and Logic Programming Languages}

\vspace*{-1mm} Constraint Logic Languages, such as SICStus Prolog,
ECLiPSe and BProlog, are extensions of Prolog and, therefore, they
are not fully-declarative. Their semantics is based on top-down
evaluation of queries (SLDNF resolution), whereas answer-set
programming is based on bottom-up evaluation.
$XSB$ is an extension of Prolog with a declarative semantics
(namely the well-founded semantics) based on top-down
evaluation of queries (OLDT resolution) with tabling.
Moreover, while answer-set languages permit \NP\ problems to be easily expressed
(it suffices to translate their logic definition into logic
programming rules), constraint logic programming languages are
procedural and the efficient implementation of \NP\ problems is
hard and time-consuming.

The relationship between ID Logic and \NPDA\ is strong since in our
language we can also distinguish components describing data, guess predicates, standard rules and constraints,
which correspond, respectively, to the ID logic components describing
data, open predicates, definitions and constraints.
Moreover, the aim of \NPDA\ is also the easy translation into different formalisms
(other than ASP), including constraint programming languages.

\subsubsection*{Answer Set Programming Languages}

\vspace*{-1mm}
%
%
The main difference of \NPDA\ with respect to DLV and Smodels
is that only restricted forms of (unstratified) negations, embedded
into built-in constructs, are allowed. As a consequence, \NPDA\ is
less expressive than DLV since the latter also uses (inclusive)
disjunction and permits expression of problems in the second level
of the polynomial hierarchy. The use of simpler languages such as
\NPDA\ allows us to avoid writing non-intuitive queries which are
difficult to optimize or translate in other formalisms for which
efficient executors exist.
It is important to observe that cardinality constraints and conditional literals of Smodels
allows us to express both subset and (generalized) partition rules as defined
in \NPDA.
This means that in Smodels it is also possible to avoid
using unstratified negation without losing expressiveness.
We also note that \emph{s-atoms} and \emph{f-atoms} of ASET-Prolog
enable us to express subset and (generalized) partition rules.

\NPDA\ is also strongly connected to Datalog  Constraint
(DC), which is also based on stable model semantics. The main
difference between \NPDA\ and DC consists in the fact that \NPDA\
forces users to write queries in a more disciplined form. In
particular, DC guesses are expressed by means of constraints (a
guess is any set of atoms in $At_C$ satisfying the constraint in
$T_C$) and there is no clear separation between $T_C$ (constraints
used to \emph{guess}) and $T_{PC}$ (constraints used to
\emph{check}). Moreover, DC only uses positive rules to infer true
atoms, whereas \NPDA\ uses stratified rules. The expressive power
of both languages captures the first level of the polynomial
hierarchy. The experiments reported in \cite{EastTru00} show that
the guess and check style of expressing hard problems can be
further optimized.

The philosophy of ASSAT is similar to that of \NPDA: while in the
ASSAT approach programs are translated in propositional logic and
then executed by means of a SAT solver, \NPDA\ programs are
translated into OPL programs and then executed by using the
ILOG OPL Development Studio. An approach similar
to the one of ASSAT is adopted by Cmodels.

A-Prolog is a general logic language 
(i.e. it allows function symbols, classical negation, head disjunction and subset rules) 
whose
semantics is based on stable models.
The aim of A-Prolog is the design of a general
language for knowledge representation and causal reasoning, whereas \NPDA\
is a simpler language (similar to the restricted FA-Prolog) which can
be easily efficiently executed and translated in other formalisms.

\section{Conclusion}

\NP\ search and optimization problems can be formulated as \DAm\
queries under non-deterministic  stable model semantics.
In order to enable a simpler and more intuitive formulation of these
problems, the \NPDA\ language has been proposed.
It is obtained by extending stratified Datalog with constraints and
two constructs for expressing partitions of relations,
so that search and optimization queries can be
expressed using only simple forms of unstratified negation.
It has also been shown that \NPDA\ captures the class
of \NP\ search and optimization problems and that \NPDA\ queries
can be easily translated into OPL programs.
An algorithm for the translation of \NPDA\ programs into OPL statements has been
provided and its correctness has been proved.
The proposed algorithm has been implemented by a system prototype which takes in input an
\NPDA\ query and gives in output an equivalent OPL program which is then
executed using the ILOG OPL Development Studio.
Consequently, \NPDA\ can also be used to define a logic interface for
constraint programming solvers.
Several experiments comparing the computation of queries by different systems
have shown the validity of our approach.

\bibliographystyle{theapa}

\end{document}